\documentclass[sigconf]{acmart}

\renewcommand\footnotetextcopyrightpermission[1]{}
\usepackage{amsmath,amsfonts}
\usepackage{algorithmic}
\usepackage{graphicx}
\usepackage{textcomp}
\usepackage{multicol}
\usepackage{multirow}
\usepackage{adjustbox}
\usepackage{enumitem}
\usepackage{booktabs}
\usepackage{hyperref}
\usepackage{makecell}
\usepackage{caption,subcaption}
\usepackage[ruled,linesnumbered]{algorithm2e}
\usepackage[normalem]{ulem}
\usepackage{pifont}

\usepackage{soul}

\graphicspath{{figures/}}

\AtBeginDocument{%
  }

\setcopyright{acmlicensed}
\copyrightyear{2018}
\acmYear{2018}
\acmDOI{XXXXXXX.XXXXXXX}

\acmConference[Conference acronym 'XX]{Make sure to enter the correct
  conference title from your rights confirmation emai}{June 03--05,
  2018}{Woodstock, NY}
\acmISBN{978-1-4503-XXXX-X/18/06}

\begin{document}


\title{Counterfactual Multi-player Bandits for Explainable Recommendation Diversification}


\author{
	Yansen Zhang$^{1}$, Bowei He{$^{1}$}, Xiaokun Zhang{$^{1}$}, Haolun Wu{$^{2}$}, Zexu Sun{$^{3}$}, {Chen Ma}$^{1\dagger}$ \\ 
	$^{1}$Department of Computer Science, City University of Hong Kong \\
	$^{2}$School of Computer Science, McGill University\\
	$^{3}$Gaoling School of Artificial Intelligence, Renmin University of China \\
	\text{yanszhang7-c@my.cityu.edu.hk}; \text{chenma@cityu.edu.hk} 
}


\renewcommand{\shortauthors}{Zhang. et al.}

\begin{abstract}
Existing recommender systems tend to prioritize items closely aligned with users' historical interactions, inevitably trapping users in the dilemma of ``filter bubble''. Recent efforts are dedicated to improving the diversity of recommendations. However, they mainly suffer from two major issues: 1) a lack of explainability, making it difficult for the system designers to understand how diverse recommendations are generated, and 2) limitations to specific metrics, with difficulty in enhancing non-differentiable diversity metrics. To this end, we propose a \textbf{C}ounterfactual \textbf{M}ulti-player \textbf{B}andits (CMB) method to deliver explainable recommendation diversification across a wide range of diversity metrics. Leveraging a counterfactual framework, our method identifies the factors influencing diversity outcomes. Meanwhile, we adopt the multi-player bandits to optimize the counterfactual optimization objective, making it adaptable to both differentiable and non-differentiable diversity metrics. Extensive experiments conducted on three real-world datasets demonstrate the applicability, effectiveness, and explainability of the proposed CMB.

\end{abstract}







\maketitle

\def\customfootnotetext#1#2{{%
		\let\thefootnote\relax
		\footnotetext[#1]{#2}}}

\customfootnotetext{2}{{$\dagger$} Corresponding author.}

\section{Introduction}
\label{sec:intro}
Recommendation systems (RS) are widely deployed in various online platforms, such as Google, Facebook, and Yahoo!, to mitigate information overload. However, most existing recommendation methods~\cite{DBLP:conf/uai/RendleFGS09,DBLP:conf/sigir/0001DWLZ020,DBLP:journals/computer/KorenBV09,cheng2016wide,zhang2021self} only prioritize recommending the most relevant items to users, which can have negative consequences for both users and service providers. Users may experience the ``filter bubble''~\cite{pariser2011filter} problem, receiving recommendations based solely on their past behavior, leading to a lack of content diversity and reduced long-term satisfaction. Content providers may face the ``Matthew Effect''~\cite{merton1968matthew}, where new or niche content/products yield less exposure, discouraging the engagement of content providers. Therefore, it is crucial to consider not only relevance but also recommendation diversity to enhance the overall user experience and maintain a healthy ecosystem for content providers.

To increase user engagement and satisfaction, as well as improve the exposure of less popular content, recommendation diversification has attracted more and more attention. Various approaches have been proposed to diversify the recommended items. Existing methods can be generally classified into three categories~\cite{wu2022survey}: pre-processing, in-processing, and post-processing methods. Pre-processing methods, like~\cite{kwon2020art,zheng2021dgcn,cheng2017learning}, intervene with the system by modifying or selecting the interaction data before the model training. In-processing methods, such as treating the need for diversity as a kind of regularization~\cite{wasilewski2016incorporating,chen2020improving} or a ranking score~\cite{li2017learning}, integrate diversification strategies into the recommendation model training process directly. Post-processing methods, such as MMR~\cite{santos2010exploiting,vargas2011intent} and DPP~\cite{gan2020enhancing,DBLP:conf/nips/ChenZZ18,gong2022dawar, huang2021sliding, wilhelm2018practical}, re-rank the recommended items based on both the item relevance scores and some diversity metrics after the model training procedure.

Unfortunately, current methods still suffer from two main limitations. Firstly, current methods, such as~\cite{DBLP:conf/nips/ChenZZ18,zheng2021dgcn,cheng2017learning,wasilewski2016incorporating,chen2020improving,vargas2011intent,gan2020enhancing,gong2022dawar,wilhelm2018practical,sanz2018enhancing}, do not provide adequate explainability regarding how factors affect the diversity of recommendations at the (latent) feature level. This limitation makes it difficult for system designers to understand the underlying drivers of diversity, hindering efforts to enhance model diversity and potentially reducing user satisfaction. Secondly, most diversification methods, like~\cite{DBLP:conf/nips/ChenZZ18,zheng2021dgcn,chen2020improving,wilhelm2018practical,li2017learning}, rely on diversity metrics to evaluate recommendation results, but they often fail to optimize these metrics directly because these metrics are mostly non-differentiable, as highlighted in a recent survey~\cite{wu2022survey}. While several methods, such as those described in~\cite{yu22aaai}, strive to optimize some diversity metrics directly, they are only suitable for very few specific non-differentiable diversity metrics, like $\alpha$-nDCG~\cite{clarke2008novelty}, and cannot handle more commonly used metrics like Prediction Coverage or Subtopic Coverage~\cite{ge2010beyond}.

To address the aforementioned challenges, we propose a counterfactual framework for explainable recommendation diversification. In response to the first limitation, we propose to identify the factors influencing diversity outcomes under the counterfactual framework. In this framework, perturbations are applied to the representation of items to adjust the diversity level of the ranking lists. Our goal is to identify the ``minimal'' changes to a specific factor in the factor space that can effectively switch the recommendation results to a desired level of diversity. Then in response to the second limitation, we design a gradient-free Counterfactual Multi-player Bandits (CMB) method to learn these perturbations by optimizing the diversity of recommended items, which is no longer constrained by diversity metrics and recommendation models. The bandit-based approach searches for the best perturbations applied to different factors, which also provides insights for explaining the recommendation diversification: \textit{the factors with more perturbations have more potential to influence both the accuracy and diversity.} Finally, as there is a growing need to achieve a better trade-off between accuracy and diversity, we redesign the optimization objective considering accuracy and diversity metrics simultaneously. Overall, our proposed approach offers a more flexible and adaptable framework that can optimize various diversity metrics directly, and provides a promising solution for explainable recommendation diversification.

To summarize, the contributions of this work are as follows:
\begin{itemize}[leftmargin=*]
    \item To explain recommendation diversification, we employ the counterfactual framework to discover the meaningful factors that affect recommendation accuracy and diversity trade-off.
    \item To optimize a range of differentiable and non-differentiable diversity metrics, we propose a bandit-based diversity optimization approach that is agnostic to diversity metrics and recommendation models.
    \item To validate the applicability, effectiveness, and explainability of our method, we conducted extensive experiments on multiple real-world datasets and diversity metrics.
\end{itemize}

\section{Related Work}
\subsection{Recommendation Diversification}

To address the ``filter bubble'' and reduced provider engagement issues, it is crucial to recommend accurate and diverse items for a healthier online marketplace. Existing diversification methods are typically offline and categorized as pre-processing, in-processing, and post-processing methods~\cite{wu2022survey}. Pre-processing methods~\cite{kwon2020art,cheng2017learning, paudel2017fewer,zheng2021dgcn} involve preparing interaction data before model training. In-processing integrates diversity into the training process, using it as regularization~\cite{chen2020improving, stella2020diversify, sanz2018enhancing, wasilewski2016incorporating, zhou2020diversifying} or a ranking score~\cite{li2017learning,yu22aaai}. Post-processing, the most scalable, includes greedy-based methods like MMR~\cite{carbonell1998use, santos2010exploiting, vargas2011intent} and DPP~\cite{gan2020enhancing,DBLP:conf/nips/ChenZZ18,gong2022dawar, huang2021sliding, wilhelm2018practical}, which adjust item selection and rankings to balance relevance and diversity, and refinement-based methods~\cite{li2020directional, tsukuda2019dualdiv}, which modify positions or replace items based on diversity metrics. Other online methods, such as bandit strategies~\cite{ding2021hybrid, li2020cascading, qin2014contextual, parapar2021diverse}, treat diversity as part of the score on each arm (item or topic) in the bandit recommendation algorithms and reinforcement learning~\cite{stamenkovic2022choosing, zheng2018drn,shi2023relieving}, continuously update based on user feedback for long-term optimization.

Although these methods enhance recommendation diversity, they do not provide explanations of the monopoly phenomenon of recommended items or the mechanisms behind their diversity improvements. Our work seeks to optimize recommendation diversity and offer explanations for these issues.

\subsection{Explainable Recommendation}

Explainable recommendations have attracted significant attention in academia and industry, aiming to enhance transparency, user satisfaction, and trust~\cite{zhang2020explainable,zhang2014explicit,zhang2025svv}. Early methods focused on generating individualized explanations, often customizing models and using auxiliary information~\cite{zhang2014explicit,lu2018coevolutionary,wu2019context}. For example, the Explicit Factor Model (EFM)~\cite{zhang2014explicit} recommends products based on features extracted from user reviews. Other approaches decouple explanations from the recommendation model, making them post-hoc and model-agnostic~\cite{ribeiro2016should, wang2018reinforcement, singh2019exs,chen2022relax}. Recently, counterfactual reasoning has been widely used to improve explainability~\cite{wachter2017counterfactual,ghazimatin2020prince,tan2021counterfactual,xiong2021counterfactual,ge2022explainable}. For instance, PRINCE~\cite{ghazimatin2020prince} identifies minimal user actions that, if removed, would change the recommendation, while CEF~\cite{ge2022explainable} uses counterfactual reasoning to explain fairness in feature-aware recommendation systems.

This work focuses on explaining recommendation diversification. While existing approaches help interpret recommendation models, they overlook diversity, which is the main focus of our work. The comparison between our method and existing diversification or explainable recommendation methods is shown in Table~\ref{tab:my_label}.

\begin{table}[t]
    \centering
    \caption{Comparisons of different recommendation methods. ``Pre'', ``In'', and ``Post'' represent Pre-processing, In-processing, and Post-processing methods, respectively. }
    \resizebox{\linewidth}{!}{
      \begin{tabular}{llcccc}
      \toprule
      \multicolumn{2}{c}{\multirow{3}{*}{Techniques}}  & \multicolumn{2}{c}{Diversity metric optimization}   & \multicolumn{2}{c}{Explanation}  \\
      \cmidrule{3-6}
      \multicolumn{2}{c}{}  & \multicolumn{1}{c}{\makecell{Differentiable\\Metric}} & \makecell{Non-differentiable \\ Metric} & \multicolumn{1}{c}{\makecell{Explanation \\ Generation}} & \makecell{Model \\ Agnostic} \\
      \midrule
      \multicolumn{1}{l}{\multirow{3}{*}{\makecell[l]{Diversified \\ Recommendation}}} & Pre~\cite{kwon2020art,cheng2017learning, paudel2017fewer,zheng2021dgcn}  & \multicolumn{1}{c}{\ding{51}}                       &             \ding{55}/\ding{51}               & \multicolumn{1}{c}{ \ding{55}}               &        \ding{55}                  \\
      \multicolumn{1}{c}{}     & In~\cite{ chen2020improving, stella2020diversify, sanz2018enhancing, wasilewski2016incorporating, zhou2020diversifying,li2017learning, yu22aaai}   & \multicolumn{1}{c}{ \ding{51}}                       &        \ding{55}/\ding{51}                     & \multicolumn{1}{c}{ \ding{55}}      &      \ding{55}        \\
      \multicolumn{1}{c}{}         & Post~\cite{DBLP:conf/nips/ChenZZ18,santos2010exploiting, vargas2011intent,gan2020enhancing, gong2022dawar, huang2021sliding, wilhelm2018practical,li2020directional, tsukuda2019dualdiv} & \multicolumn{1}{c}{ \ding{51}}                       &               \ding{55}/\ding{51}     & \multicolumn{1}{c}{ \ding{55}}               &          \ding{55}                \\
      \midrule
      \multicolumn{1}{l}{\makecell[l]{Explainable \\ Recommendation}}     & \multicolumn{1}{c}{\makecell[l]{\cite{ge2022explainable,ghazimatin2020prince,lu2018coevolutionary,ribeiro2016should,singh2019exs,tan2021counterfactual} \\ \cite{wachter2017counterfactual,wang2018reinforcement,wu2019context,xiong2021counterfactual,zhang2014explicit}}}
      & \multicolumn{1}{c}{ \ding{55}}                       &             \ding{55}                & \multicolumn{1}{c}{ \ding{51}}               &            \ding{51}              \\
      \midrule
      \multicolumn{2}{l}{\textbf{CMB (ours)}}                                                    & \multicolumn{1}{c}{ \ding{51}}                       &             \ding{51}                & \multicolumn{1}{c}{ \ding{51}}               &          \ding{51}                \\
      \bottomrule
    \end{tabular}
    }
    \label{tab:my_label}
    \vspace{-2.5mm}
\end{table}

\section{Methodology}
\subsection{Preliminaries}
We first formulate the problem of explainable recommendation diversification. Then, we introduce two base recommendation models, as well as several representative diversity metrics.

\subsubsection{\textbf{Problem Formulation}}

Given a user set $\mathcal{U}$, an item set $\mathcal{V}$, and the corresponding user-item interactions set $\mathcal{T}$, the purpose of explainable diversification is to recommend accurate and also diverse items that meet user interests, while offering explainability to the diversification. Formally, we need to provide the diverse top-$K$ recommendation list $R^{u} \subset \mathcal{V} (\vert R^{u} \vert = K)$ to each user $u$, and analyze what leads to the diversified results.

\subsubsection{\textbf{Base Recommendation Models}}
\label{base-model}
Given the user latent feature matrix $\mathbf{P} \in \mathbb{R}^{d \times |\mathcal{U}|}$ and item latent feature matrix $\mathbf{Q} \in \mathbb{R}^{d \times |\mathcal{V}|}$, where $d$ is the dimension of the latent feature matrices. We define a base recommendation model $g$ that predicts the user-item ranking score $\hat{y}_{u,v}$ for user $u$ and item $v$ by:
\begin{equation}
    \hat{y}_{u,v} = g(\mathbf{p}_u, \mathbf{q}_v \mid \mathbf{Z}, \mathbf{\Theta}),
\end{equation}
where $\mathbf{p}_u \in \mathbb{R}^{d}$ and $\mathbf{q}_v \in \mathbb{R}^{d}$ are the latent feature vector of user $u$ and item $v$, respectively.
The symbol $\mathbf{\Theta}$ denotes the model parameters,
and $\mathbf{Z}$ represents all other auxiliary information.
Since collaborative filtering (CF) methods are still mainstream in current systems, we mostly work on the latent features of CF methods.
Without loss of generality, we can also target the raw features (e.g., age, gender, etc.), which will be discussed in Sec.~\ref{exp:case study}. 
We explore two popular and effective instances of $g$ as follows~\cite{ye2021dynamic,yang2023dgrec}.

\begin{itemize}[leftmargin=*]
    \item \textbf{BPRMF}~\cite{DBLP:conf/uai/RendleFGS09}. The prediction score of BPRMF is as follows:
          \begin{equation}
              \hat{y}_{u,v} = \mathbf{p}_{u}^{\top} \cdot {\mathbf{q}_v},
          \end{equation}
          where $\cdot$ denotes the dot product. Then the Bayesian Personalized Ranking loss function is adopted:
          \begin{equation}
              \mathcal{L}_{bpr} = \underset{(u, i, j)}{\sum} -ln \sigma (\hat{y}_{u,i} - \hat{y}_{u,j}) + \lambda_{\mathbf{\Theta}} \Vert \mathbf{\Theta} \Vert^2,
              \label{equ:bpr}
          \end{equation}
          where $(u, i, j)$ is the user-item triplet in which $j$ is sampled from the non-interacted items of user $u$. The symbol $\lambda_{\mathbf{\Theta}}$ indicates the regularization hyperparameter.

    \item \textbf{LightGCN}~\cite{DBLP:conf/sigir/0001DWLZ020}. The LightGCN model consists of $L$ graph convolution network (GCN) layers. The graph convolution operation in each layer $l$ is defined as:
          \begin{equation}
              \begin{aligned}
                  \mathbf{p}^{(l)}_u = \underset{v \in \mathcal{N}_u}{\sum} \frac{1}{\sqrt{\vert \mathcal{N}_u \vert \vert \mathcal{N}_v \vert}} \mathbf{q}^{(l-1)}_v, \,
                  \mathbf{q}^{(l)}_v = \underset{u \in \mathcal{N}_v}{\sum} \frac{1}{\sqrt{\vert \mathcal{N}_v \vert \vert \mathcal{N}_u \vert}} \mathbf{p}^{(l-1)}_u,
              \end{aligned}
          \end{equation}
          where $\mathbf{p}^{(0)}_u = \mathbf{p}_u, \mathbf{q}^{(0)}_v = \mathbf{q}_v$, and $\mathcal{N}_u$ ($\mathcal{N}_v$) denotes the set of items (users) that are interacted with user $u$ (item $v$). The model prediction score is formulated as follows:
          \begin{equation}
              \hat{y}_{u,v} = {\left(\frac{1}{L+1} \underset{l=0}{\sum^L}\mathbf{p}^{(l)}_u\right)}^{\mathrm{\top}} \cdot \left(\frac{1}{L+1} \underset{l=0}{\sum^L} \mathbf{q}^{(l)}_v\right).
          \end{equation}
          When utilizing LightGCN as the base recommendation model, the training loss is consistent with the BPRMF.
\end{itemize}

\subsubsection{\textbf{Diversity Metrics}}
\label{sec:metric}
Among all diversity metrics, we discuss the following four most popular metrics~\cite{kunaver2017diversity,cheng2017learning,han2019geographic,wu2022survey}.

\begin{itemize}[leftmargin=*]
    \item \textbf{Novelty-biased Normalized Discounted Cumulative Gain ($\alpha$-nDCG)}~\cite{clarke2008novelty}.
          The $\alpha$-nDCG is a subtopic-level metric based on Normalized Discounted Cumulative Gain (NDCG) but considers the subtopic and is redundancy-aware along the items. The hyperparameter $\alpha$ is a geometric penalization for redundancy.

    \item \textbf{Subtopic Coverage (SC)}~\cite{ge2010beyond, he2019diversity}. SC@K is a subtopic-level coverage of a recommended item list $R^u$ in the whole item set.

    \item \textbf{Prediction Coverage (PC)}~\cite{ge2010beyond, herlocker2004evaluating}. PC@K is an item-level coverage of all recommendation lists $R^{u}$ in the whole item set.

    \item \textbf{Intra-List Average Distance (ILAD)}~\cite{zhang2008avoiding}.
          ILAD@K is an item-level distance-based metric and evaluates the diversity by calculating the average dissimilarity of all pairs of items in the recommendation list $R^u$. In this work, we adopt cosine similarity to calculate the dissimilarity as the distance function.
\end{itemize}

\subsection{Counterfactual Framework for Explainable Diversification}
Current diversification approaches generate diverse lists that are hard to explain and control. However, understanding the underlying diversity mechanism is crucial for making intelligent decisions in real-world applications.
Inspired by counterfactual reasoning~\cite{wachter2017counterfactual,ge2022explainable}, we develop a perturbation-based framework for explaining the diversification of the recommendation lists. 

The essential idea behind the proposed explanation model is to discover a perturbation matrix $\Delta$ on items' features by solving a counterfactual optimization problem that maximizes diversity, as well as identify which features are the underlying drive of diversified recommendations. After identifying these features, it is easy to generate feature-based explanations for the given recommendation model $g$ and guide the system to make appropriate decisions that increase the recommendation diversity. Generally, given a recommendation model $g$, we have a certain recommendation result $\mathcal{R}_g = \{R^{u_1}, R^{u_2}, \cdots, R^{u_i}, \cdots, R^{u_{|\mathcal{U}|}}\} (|R^{u_i}| = K, i=1,2,\cdots,|\mathcal{U}|)$ containing all users' top-$K$ recommendation lists, where $R^{u_i}$ represents the top-$K$ items list recommended to user $u_i$ by the base model $g$. We denote the recommendation diversity of $g$ as,
\begin{equation}
    \Psi=Diversity(\mathcal{R}_g),
\end{equation}
where $Diversity(\cdot)$ can be any of the previously introduced diversity measurements in Sec.~\ref{sec:metric}. 

Specifically, for the learned item latent feature matrix $\mathbf{Q} \in \mathbb{R}^{d \times |\mathcal{V}|}$ from $g$, we slightly intervene with an equal-size matrix $\Delta \in \mathbb{R}^{d \times |\mathcal{V}|}$. In detail, a small perturbation $\Delta_{i,j}$  will be added to feature $i$ of item $j$ ($\mathbf{Q}_{i,j}$) to obtained the perturbed input $\widetilde{\mathbf{Q}}$. That is,
\begin{equation}
    \widetilde{\mathbf{Q}} = \mathbf{Q} + \Delta.
\end{equation}
With this perturbed item latent feature matrix $\widetilde{\mathbf{Q}}$, the base model $g$ will change the recommendation from $\mathcal{R}_g$ to a new counterfactual result $\widetilde{\mathcal{R}}_g$ with a new diversity measure $\widetilde{\Psi}$,
\begin{equation}
    \widetilde{\Psi}=Diversity(\widetilde{\mathcal{R}}_g).
\end{equation}
Here, our goal is to find the minimum intervention on item features that will result in the maximum improvement in terms of diversity. Thus, the objective function would be:
\begin{equation}
    \mathop{\max}_{\Delta} \| \widetilde{\Psi} \|^2_2 - \lambda_1 \|\Delta\|_{1},
    \label{eqn:obj}
\end{equation}
where $\lambda_{1}$ is a hyperparameter that controls the balance between two terms: the first maximizes the predefined diversity, and the second constrains the perturbation by reflecting the distance between the original input and the counterfactuals. To minimize changes in item latent features, we apply $L_1$ norm constraint on $\Delta$ and scale its absolute values between $[0, 1]$, encouraging more $\Delta$ as $0$ and highlighting the features that most influence diversity.

\begin{figure}
    \centering
    \includegraphics[width=\linewidth]{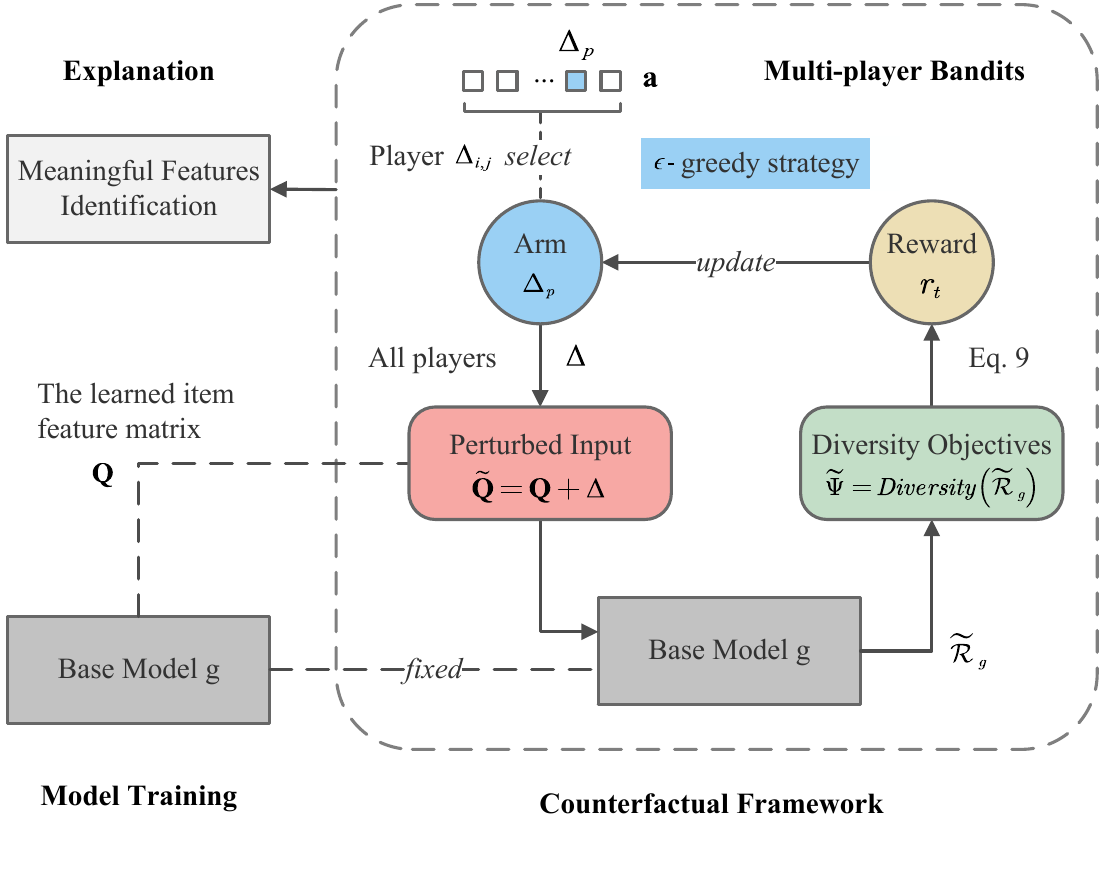}
    \setlength{\abovecaptionskip}{-5pt}
    \caption{The architecture of CMB. CMB consists of three major stages: the first stage of base model training, the second stage of counterfactual framework within multi-player bandits optimization, and the third stage of explanation.
	The $\mathbf{Q}$ is the learned item latent feature matrix from the base model $g$.
	The $\widetilde{\Psi}$ in the green part is alterable, which can be diversity or the trade-off between diversity and accuracy.}
    \Description{The architecture of CMB. CMB consists of three major stages: the first stage of base model training, the second stage of counterfactual framework within multi-player bandits optimization, and the third stage of explanation.}
    \label{cmb}
\end{figure}

\subsection{Multi-player Bandits for Diversity Optimization}

According to certain needs of the diversity, various metrics can be used to optimize Eq.~\ref{eqn:obj} for learning the perturbation $\Delta$. For example, $\alpha$-nDCG or SC metric can be employed to ensure broader coverage of subtopics in the recommendation list, while ILAD or PC metric can be used to enhance item-level diversity.

However, a significant challenge is that most diversity metrics are non-differentiable, making it difficult to define a proxy for their optimization~\cite{wu2022survey,yu22aaai}. For instance, among the four metrics discussed in Sec.~\ref{sec:metric}, only ILAD is differentiable, complicating the integration of non-differentiable metrics into gradient-based counterfactual frameworks. To overcome this, we propose a bandit-based method to learn the perturbation matrix $\Delta$, enabling the optimization of diverse objectives without relying on gradient computation.



The multi-armed bandit problem~\cite{bubeck2009pure, karnin2013almost, umami2021comparing} models decision-making under uncertain rewards, where a player chooses among various options (``arms'') to maximize cumulative payoff by balancing exploration and exploitation. This approach is well-suited for optimizing non-differentiable objectives within a counterfactual framework. Therefore, in our specific problem, each item feature is treated as a player, selecting an appropriate arm to construct the final perturbation matrix. Moreover, our problem can be further conceptualized as a multi-player bandit scenario~\cite{bistritz2020cooperative, bistritz2018distributed, bubeck2021cooperative}, where players collaborate to optimize a shared reward, which serves as the objective in the counterfactual framework.

Specifically, we treat each variable $\Delta_{i,j}$ in $\Delta$ as a player; for simplicity, we denote it as $p (|p| = d \times |\mathcal{V}|)$, and then select a suitable arm $\Delta_p$ from arms $\mathbf{a}$ for every $p$ iteratively to maximize the objective in the counterfactual framework. Assume the total number of iterations for one player $p$ to select an arm is $T$; our problem can be formulated by maximizing the following $T$-step cumulative reward for each player:
\begin{equation}
    V^T =  \max \sum_{t=1}^{T} r_t,
\end{equation}
where $r_t$ is the reward at the iteration step $t$ obtained from all players by selecting arms, as calculated by Eq.~\ref{eqn:obj}. In each iteration, every player selects a particular arm and gets $\Delta$ from all players to calculate the reward. This process continues as players select arms in subsequent iterations based on the obtained rewards. The iterative process repeats until the final iteration, where each player selects the optimal arm to achieve diversification.
At this point, the value of the perturbation $\Delta$ can be justified.

More specifically, before commencing the algorithm, it is necessary to define the arm values from which the players can make their selections. To achieve this, we utilize the following method to initialize the arms $\mathbf{a}$ for each player $p$:
\begin{equation}
    \mathbf{a} = \mathtt{INIT}(A, n_A),
\end{equation}
where $A$ and $n_A$ represent the perturbation threshold and the number of arms, respectively. The $\mathtt{INIT}(\cdot)$ method samples $n_A$ values evenly from $[-A, A]$. For simplicity, this initialization is applied to each player, and the item latent feature matrix $\mathbf{Q}$ is scaled to $[-1, 1]$ using maximum absolute scaling. At each iteration $t$, players independently choose an arm $\Delta_p$ from $\mathbf{a}$ based on action selection strategies like $\epsilon$-greedy or UCB~\cite{li2010contextual,bubeck2012regret,kaufmann2012bayesian,slivkins2019introduction}, with experiments showing that $\epsilon$-greedy is more efficient and effective. The selection strategy of $\epsilon$-greedy strategy is as follows:
\begin{equation}
    \Delta_{p}^{t} = \begin{cases}
        \mathop{\arg\max}_{\Delta_p \in \mathbf{a}}(V_{\mathbf{a}}^{t}) & \text{with probability } 1-\epsilon, \\
        \text{a random arm}                                             & \text{with probability } \epsilon,
    \end{cases}
\end{equation}
where $\Delta_p^t$ and $V_{\mathbf{a}}^{t}$ are the arm value selected by the player and the cumulative reward vector containing all arms for the player in the $t$ iteration, respectively.

When all players have selected the arm $\Delta_p^t$
, we can get the $\Delta$ and further the reward $r_t$ (Eq.~\ref{eqn:obj}) based on the counterfactual result $\widetilde{\mathcal{R}}_g$. To reduce the computational complexity,
we update the cumulative reward value $V^{t+1}_{\Delta_p}$ of the corresponding arm $\Delta_p$ selected by each player $p$ in iteration $t+1$ by an incremental average method:
\begin{align}
	V^{t+1}_{\Delta_p} &= \frac{1}{n}\sum_{i=1}^{n} r_i, \nonumber\\
	&= \frac{1}{n} \left(r_n^t + \sum_{i=1}^{n-1} r_i\right), \nonumber\\
	&= \frac{1}{n} \left(r_t + \left(n-1\right)\frac{1}{n-1} \sum_{i=1}^{n-1} r_i\right), \nonumber\\
	&= \frac{1}{n} \left(r_t + \left(n-1\right)V^{t}_{\Delta_p}\right), \nonumber\\
	&= \frac{1}{n} \left(r_t + n V^{t}_{\Delta_p} - V^{t}_{\Delta_p}\right), \nonumber\\
	& = V^{t}_{\Delta_p} + \frac{1}{n} \left(r_t - V^{t}_{\Delta_p}\right),
	\label{equ:arm value}
\end{align}
where $n$ is the times that arm $\Delta_p$ has been selected by player $p$ till iteration $t$. Thus, the model iteratively learns and adjusts the $\Delta$ until convergence. In order to simplify the problem, we assume that different players are independent of each other.
The details of the CMB algorithm are described in Algorithm~\ref{algorithm}.

The challenge in diversified recommendation is to enhance diversity while preserving accuracy, i.e., maximizing diversity without significantly compromising accuracy. To better balance these two aspects, we propose a redesign of the optimization objective (Eq.~\ref{eqn:obj}), particularly the counterfactual diversity measurement $\widetilde{\Psi}$. As previously discussed, while a suitable diversity metric for $\widetilde{\Psi}$ can be chosen, it often leads to some loss in accuracy. To achieve an optimal balance, inspired by~\cite{carbonell1998use,DBLP:conf/nips/ChenZZ18}, we redesign $\widetilde{\Psi}$ to balance both accuracy and diversity simultaneously. Specifically,
\begin{equation}\label{eq2}
    \widetilde{\Psi} = \lambda_2 \times Accuracy(\widetilde{\mathcal{R}}_g) + (1 - \lambda_2) \times Diversity(\widetilde{\mathcal{R}}_g),
\end{equation}
where $\lambda_2$ is a hyperparameter to control the trade-off between accuracy and diversity, and $Accuracy(\cdot)$ and $Diversity(\cdot)$ represent accuracy metrics (e.g., Recall@K, NDCG@K, etc.) and diversity metrics (Sec.~\ref{sec:metric}), respectively.

\noindent
\textbf{Discussion}. (1) We mainly work on the latent features of items to explain the recommendation diversification. The key consideration is that current mainstream recommendation models still originate from collaborative filtering methods, which are based on latent features. These motivate us to work on the latent features for controlling the diversity level of recommendation results. It is worth noting that our model can both work raw and latent features.
(2)
The main reason why we decided to apply the perturbation $\Delta$ to the features of items is that perturbing at the items' feature level enables us to identify specific features that impact the diversity of the model at a more fine-grained feature level. (3) The multi-armed bandits method, unlike previous approaches, is primarily utilized in online recommendation scenarios. In our study, we employ this method to optimize the learning objective in the counterfactual framework, especially targeting and optimizing the non-differentiable diversity metrics directly.

\subsection{Meaningful Features Identification as Explanation}
\label{explainabtion}

Once finishing optimization, we get the ``minimal'' changes $\Delta$ and the corresponding recommendation results under such changes. The values of $\Delta$ indicate the influence of item features on the accuracy-diversity trade-off of the recommendation lists generated by the base model $g$. Specifically, compared with the initial item latent feature matrix $\mathbf{Q}$, after adding the values of $\Delta$, the model $g$ is supposed to generate more diverse lists. Therefore, the perturbation $\Delta$ provides insights for our explanation. In particular, larger absolute values of $\Delta$ correspond to a greater need for the corresponding features to promote greater diversity.

Based on the above analysis, after we identify each feature's ``ability'' to incur the diversity of the recommendation list, we further select the most meaningful features of the items affecting diversity and give insights into recommendation systems. We provide two perspectives on detecting the most meaningful features here, namely CMB-Individual and CMB-Shared.
\begin{equation}
    \begin{cases}
        \text{CMB-Individual}, & \textit{feature-level}        \\
        \text{CMB-Shared},     & \textit{item-level} \nonumber
    \end{cases}
\end{equation}
Specifically, the strategy of CMB-Individual is to directly select the features corresponding to the higher absolute values of $\Delta$ on the features as an explanation for each user.
The CMB-Shared strategy is that we take the absolute values of $\Delta$, compute the mean value by rows, and compress the $\Delta$ into a vector $\Delta_v \in \mathbb{R}^d$,
\begin{equation}
    \Delta_v = MEAN(|\Delta|, dim=0),
    \label{eqn:delta}
\end{equation}
and then choose the features corresponding to the higher values of $\Delta_v$ as an explanation. After discovering the most meaningful features, we can adjust the values of these features to meet the corresponding needs of diversity.

\textbf{\begin{algorithm}
        \caption{The CMB algorithm.}
        \label{algorithm}
        \KwIn{The base model $g$, $\epsilon$, $A$, $n_A$.}
        \KwOut{The learned $\Delta$.}
        Initialization: \\
        \For{player p = $1,...,d \times |\mathcal{V}|$}{
            $\Delta_{p}$ = 0\;
            Sample arms for each player $p$ with $\mathbf{a} = numpy.linspace(-A, A, n_A)$\;
            Initialize cumulative reward value $V_{\mathbf{a}}$ = 0 for each arm in $\mathbf{a}$\;
        }
        Optimization: \\
        \For{t = $1, ..., T$}{
            \For{player p = $1, ..., d \times |\mathcal{V}|$}{
                Sample $random$ from uniform distribution $U(0, 1)$\;
                \eIf{random $<$ $\epsilon$}{
                    $\Delta_{p}^t$ = select an arm at random\;
                }{
                    $\Delta_{p}^t$ = select the arm which has largest $V_{\mathbf{a}}^t$\;
                }
            }
            $\Delta$ = the combination of the $\Delta_{p}$ of each player\;
            Add $\Delta$ to the item latent feature matrix $\mathbf{Q}$ and get perturbed item latent feature matrix $\widetilde{\mathbf{Q}}$\;
            Obtain the $\widetilde{\Psi}$ according to counterfactual recommendation result $\widetilde{\mathcal{R}}_g$\;
            Receive the feedback $r_t$ and update the values of selected arms with Eq.~\ref{equ:arm value}\;
        }
    \end{algorithm}
}

\subsection{Overall Procedure}

The entire procedure contains three stages, as shown in Fig.~\ref{cmb}. The base model $g$ introduced above will be trained in the first stage. Then, in the second stage, the counterfactual framework is first constructed. Based on this, the bandit algorithm is used to learn the perturbations by optimizing the diversity of the recommended top-$K$ lists. Our framework is model-agnostic and applicable to any recommendation model $g$. Meanwhile, our model is metric-agnostic since the optimization objective can be any diversity metric. In the final stage, two strategies are utilized to discover the most meaningful features for recommendation diversification.

\section{Experiments}
In this section, we conducted experiments on three real-world datasets to demonstrate the effectiveness of our method. We mainly focus on the following questions:

\begin{itemize}[leftmargin=*]
	\item \textbf{RQ1:} Does our method get better recommendation diversification effects than the state-of-the-art methods? Especially the trade-off between recommendation accuracy and diversity.
	\item \textbf{RQ2:} Do the selected top features play a significant role in diversification performance?


	\item \textbf{RQ3:} Are our generated feature-level diversification explanations reasonable and intuitive in real cases?
	\item \textbf{RQ4:} Do the hyperparameters ($A$, $n_A$, etc.) play a significant role in diversification performance?
\end{itemize}
\subsection{Experiment Setup}
\subsubsection{\textbf{Datasets}}
We performed experiments on three widely used real-world datasets, \textit{MovieLens 1M}~\cite{harper2015movielens} (\textit{ML1M}), \textit{MovieLens 10M}~\cite{harper2015movielens} (\textit{ML10M}), and \textit{Amazon CDs and Vinyl}~\cite{DBLP:conf/emnlp/NiLM19} (\textit{CDs}) to evaluate the models under different data scales, data sparsity, and application scenarios. The statistics of the datasets are shown in Table~\ref{dataset}.


For all datasets, we convert ratings to implicit feedback, treating ratings no less than four (out of five) as positive and all other ratings as missing entries. To optimize base models, we randomly sample 3 negative instances for each user's positive interaction. For the \textit{CDs} dataset, we use the top 30 categories with the highest frequency as subtopic information according to the metadata. Each dataset is split 8:1:1 for training, test, and validation. We independently run all models five times and report the average results.

\subsubsection{\textbf{Baselines}}
To verify the effectiveness of our method, we compare it with the following representative baselines. Two vanilla recommendation models \textbf{BPRMF}~\cite{DBLP:conf/uai/RendleFGS09} and \textbf{LightGCN}~\cite{DBLP:conf/sigir/0001DWLZ020}, which are introduced in Sec.~\ref{base-model}. Two recommendation diversification methods \textbf{MMR}~\cite{carbonell1998use} and \textbf{DPP}~\cite{DBLP:conf/nips/ChenZZ18}. In addition, we also explore the method \textbf{$\text{CMB}^{\text{Gradient}}$}, which represents our method is directly optimized for differentiable metrics (e.g., ILAD) by using the gradient method instead of the bandit method.

\subsubsection{\textbf{Evaluation Metrics}}
In all experiments, we evaluate the recommendation performance using accuracy and diversity metrics on Top-$K$ ($K$ = {10, 20}) lists, where $K$ represents the list length as discussed earlier during training. For all metrics, the higher the value is, the better the performance is.

\textbf{Accuracy}. We evaluate the accuracy of the ranking list using Recall@K and NDCG@K.

\textbf{Diversity}. We evaluate the recommendation diversity by the $\alpha$-nDCG@K, SC@K, PC@K, and ILAD@K introduced previously.

\subsubsection{\textbf{Implementation Details}}
In our experiments, we employ BPRMF and LightGCN as the base model $g$, with LightGCN set to three layers of the graph neural network. For baselines, BPRMF calculates the relevance scores for MMR and constructs the kernel matrix for DPP. The trade-off parameter for MMR and DPP is empirically set to 0.9 after testing values from \{0.1, 0.3, 0.5, 0.7, 0.9\}, prioritizing guaranteeing the optimal accuracy performance as much as possible. All models have a latent feature dimension of 50, with $K$ set to 20 for counterfactual learning. The $\alpha$ in $\alpha$-nDCG is set to 0.5 same as~\cite{santos2010exploiting,vargas2011intent,li2017learning,qin2020diversifying,yu22aaai}, and model parameters are optimized by Adam~\cite{kingma2014adam} with a learning rate of 0.005. In the bandit algorithm, we set the threshold of arm values $A$ to 0.3, the number of arms $n_A$ to 61, and $\epsilon$ to 0.1. The hyperparameters $\lambda_{1}$, $\lambda_2$, and the total number of iterations $T$ are set to 5, 0.9, and 200, respectively.
The source code is available at \url{https://github.com/Forrest-Stone/CMB}.
\begin{table}
  \caption{The statistics of datasets.}
  \label{dataset}
  \resizebox{\linewidth}{!}{
  \begin{tabular}{lrrrrr}
  \toprule
  Dataset & \#User & \#Item & \#Subtopic & \#Interaction & Density \\
  \midrule
  \textit{ML1M} & 5,950 & 3,125 & 18 & 573,726 & 0.0309 \\
  \textit{ML10M} & 51,692 & 7,135 & 19 & 4,752,578 & 0.0129 \\
  \textit{CDs} & 13,364 & 29,294 & 30 & 371,204 & 0.0009 \\
  \bottomrule
  \end{tabular}
  }
\end{table}

\begin{table*}
    \caption{Comparisons of the accuracy and diversity performance. The base model $g$ here adopts BPRMF. The bold scores are the best in each column, and the underlined scores are the second best. The symbols $\uparrow$ and $\downarrow$, along with their preceding values, represent the percentage (\% is omitted) improvement and decrease of a given method in the corresponding metric, in comparison to the base model $g$.} 
    \label{rq11}
    \centering
    \begin{adjustbox}{max width=\textwidth}
    \begin{tabular}{lllllll}
    \toprule
    Metric
    & Recall@10
    & NDCG@10
    & $\alpha$-nDCG@10
    & SC@10
    & PC@10
    & ILAD@10 \\
    \midrule
    \multicolumn{7}{c}{\textit{ML1M}} \\
    \midrule
        BPRMF & \textbf{0.1465} & \textbf{0.2742} & 0.7035 & 0.4993 & 0.3206 & 0.2010 \\
        \midrule
        MMR & 0.0441 (69.90$\downarrow$) & 0.0741 (72.98$\downarrow$) & 0.6980 (0.78$\downarrow$) & 0.4692 (6.03$\downarrow$) & 0.0970 (69.74$\downarrow$) & 0.1709 (14.98$\downarrow$) \\
        DPP & 0.0360 (75.43$\downarrow$) & 0.0689 (74.87$\downarrow$) & \textbf{0.7186} (2.15$\uparrow$) & \textbf{0.5558} (11.32$\uparrow$) & \textbf{0.4554} (42.05$\uparrow$) & \textbf{0.3693} (83.73$\uparrow$) \\
        $\text{CMB}_{\text{BPRMF}}$-$\alpha$-nDCG-Recall & 0.1388 (5.26$\downarrow$) & 0.2588 (5.62$\downarrow$) & 0.7094 (0.84$\uparrow$) & 0.5097 (2.08$\uparrow$) & 0.3291 (2.65$\uparrow$) & \underline{0.2253} (12.09$\uparrow$) \\
        $\text{CMB}_{\text{BPRMF}}$-SC-Recall & \underline{0.1391} (5.05$\downarrow$) & \underline{0.2594} (5.40$\downarrow$) & 0.7078 (0.61$\uparrow$) & 0.5102 (2.18$\uparrow$) & 0.3307 (3.15$\uparrow$) & 0.2246 (11.74$\uparrow$) \\
        $\text{CMB}_{\text{BPRMF}}$-PC-NDCG & 0.1388 (5.26$\downarrow$) & 0.2584 (5.76$\downarrow$) & 0.7093 (0.82$\uparrow$) & 0.5107 (2.28$\uparrow$) & 0.3286 (2.50$\uparrow$) & 0.2247 (11.79$\uparrow$) \\
        $\text{CMB}_{\text{BPRMF}}$-ILAD-NDCG & 0.1387 (5.32$\downarrow$) & 0.2587 (5.65$\downarrow$) & \underline{0.7109} (1.05$\uparrow$) & \underline{0.5127} (2.68$\uparrow$) & \underline{0.3313} (3.34$\uparrow$) & 0.2246 (11.74$\uparrow$) \\
        \midrule
    \multicolumn{7}{c}{\textit{ML10M}} \\
    \midrule
        BPRMF & \textbf{0.1549} & \textbf{0.2648} &  0.7043 & 0.5483 & 0.2453 & 0.1886 \\
        \midrule
        MMR & 0.0402 (74.05$\downarrow$) & 0.0602 (77.27$\downarrow$) & 0.7095 (0.74$\uparrow$) & 0.5155 (5.98$\downarrow$) & 0.0416 (83.04$\downarrow$) & 0.1623 (13.94$\downarrow$) \\
        DPP & 0.0253 (83.67$\downarrow$) & 0.0541 (79.57$\downarrow$) & 0.6977 (0.94$\downarrow$) & \textbf{0.6072} (10.74$\uparrow$) & \textbf{0.3735} (52.26$\uparrow$) & \textbf{0.3764} (99.58$\uparrow$) \\
        $\text{CMB}_{\text{BPRMF}}$-$\alpha$-nDCG-Recall & 0.1488 (3.94$\downarrow$) & 0.2531 (4.42$\downarrow$) & \textbf{0.7126} (1.18$\uparrow$) & 0.5544 (1.11$\uparrow$) & 0.2475 (0.90$\uparrow$) & 0.2045 (8.43$\uparrow$) \\
        $\text{CMB}_{\text{BPRMF}}$-SC-Recall & \underline{0.1490} (3.81$\downarrow$) & \underline{0.2535} (4.27$\downarrow$) & \underline{0.7123} (1.14$\uparrow$) & 0.5544 (1.11$\uparrow$) & \underline{0.2477} (0.98$\uparrow$) & 0.2046 (8.48$\uparrow$) \\
        $\text{CMB}_{\text{BPRMF}}$-PC-NDCG & 0.1487 (4.00$\downarrow$) & 0.2532 (4.38$\downarrow$) & \underline{0.7123} (1.14$\uparrow$) & \underline{0.5546} (1.15$\uparrow$) & 0.2471 (0.73$\uparrow$) & \underline{0.2051} (8.75$\uparrow$) \\
        $\text{CMB}_{\text{BPRMF}}$-ILAD-NDCG & 0.1486 (4.07$\downarrow$) & 0.2527 (4.57$\downarrow$) & 0.7110 (0.95$\uparrow$) & 0.5541 (1.06$\uparrow$) & 0.2460 (0.29$\uparrow$) & 0.2050 (8.70$\uparrow$) \\
    \midrule
    \multicolumn{7}{c}{\textit{CDs}} \\
    \midrule
        BPRMF & \textbf{0.0515} & \textbf{0.0457} & \underline{0.7206}& 0.1700 & 0.1665 & 0.2332 \\
        \midrule
        MMR & 0.0033 (93.59$\downarrow$) & 0.0032 (93.00$\downarrow$) & \textbf{0.7240} (0.47$\uparrow$) & 0.1705 (0.29$\uparrow$) & 0.0247 (85.17$\downarrow$) & 0.2372 (1.72$\uparrow$) \\
        DPP & 0.0115 (77.67$\downarrow$) & 0.0128 (71.99$\downarrow$) & 0.7116 (1.25$\downarrow$) & \textbf{0.2409} (41.71$\uparrow$) & \textbf{0.3261} (95.86$\uparrow$) & \textbf{0.4013} (72.08$\uparrow$) \\
        $\text{CMB}_{\text{BPRMF}}$-$\alpha$-nDCG-NDCG & \underline{0.0477} (7.38$\downarrow$) & \underline{0.0422} (7.66$\downarrow$) & 0.7183 (0.32$\downarrow$) & \underline{0.1739} (2.29$\uparrow$) & \underline{0.1825} (9.61$\uparrow$) & 0.2511 (7.68$\uparrow$) \\
        $\text{CMB}_{\text{BPRMF}}$-SC-NDCG & 0.0475 (7.77$\downarrow$) & 0.0421 (7.88$\downarrow$) & 0.7192 (0.19$\downarrow$) & 0.1736 (2.12$\uparrow$) & 0.1824 (9.55$\uparrow$) & 0.2510 (7.63$\uparrow$) \\
        $\text{CMB}_{\text{BPRMF}}$-PC-Recall & 0.0476 (7.57$\downarrow$) & 0.0421 (7.88$\downarrow$) & 0.7180 (0.36$\downarrow$) & 0.1737 (2.18$\uparrow$) & 0.1816 (9.07$\uparrow$) & 0.2509 (7.59$\uparrow$) \\
        $\text{CMB}_{\text{BPRMF}}$-ILAD-Recall & \underline{0.0477} (7.38$\downarrow$) & \underline{0.0422} (7.66$\downarrow$) & 0.7189 (0.24$\downarrow$) & 0.1736 (2.12$\uparrow$) & 0.1823 (9.49$\uparrow$) & \underline{0.2513} (7.76$\uparrow$) \\
    \bottomrule
    \end{tabular}
    \end{adjustbox}
    \end{table*}

\subsection{Performance Comparison (RQ1)}
Tables~\ref{rq11}, \ref{rq12}, and \ref{rq13} compare the performance of different methods under two base models. Our model is denoted as $\text{CMB}_{\text{BPRMF}}$ when using BPRMF, and $\text{CMB}_{\text{LightGCN}}$ when using LightGCN. The * in CMB-* represents the specific objective $\widetilde{\Psi}$ optimized in Eq.~\ref{eqn:obj}. For instance, CMB-$\alpha$-nDCG means that we adopt the $\alpha$-nDCG metric for $\widetilde{\Psi}$, while CMB-$\alpha$-nDCG-NDCG means that we use the trade-off optimization objective (Eq.~\ref{eq2}) of $\alpha$-nDCG and NDCG. When optimizing a single diversity metric, $\lambda_{1}$ is set to 0 for optimal results. BPRMF is the default base model $g$ unless otherwise specified. The observations from Tables~\ref{rq11}, \ref{rq12}, and \ref{rq13} are as follows.

\noindent \textbf{Trade-off observations}.
First, as shown in Table~\ref{rq11}, CMB achieves a better balance between accuracy and diversity than other methods. While diversification methods like DPP improve diversity performance, they often significantly compromise accuracy performance. For example, DPP increases ILAD@10 on \textit{ML1M} by 83.73\% but causes a 75.43\% drop in Recall@10, which is counterproductive to the primary goal of recommendation systems. In contrast, our method achieves an acceptable balance between accuracy and diversity. For instance, $\text{CMB}_{\text{BPRMF}}$-SC-Recall increases ILAD@10 by 11.74\% on \textit{ML1M} and 8.48\% on \textit{ML10M}, with only a 5.05\% and 3.81\% reduction in Recall@10, respectively. Similar trends are observed in other CMB variants, as evidenced in Tables~\ref{rq11} and \ref{rq12}, demonstrating the effectiveness of our trade-off objective.

Second, not only does the combined optimization objective help achieve a reasonable balance between accuracy and diversity, but also the single diversity objective does. For instance, as shown in Table~\ref{rq13}, when compared to BPRMF, $\text{CMB}_{\text{BPRMF}}$-ILAD shows a decrease of 18.79\% in Recall@10 and an increase of 40.62\% in ILAD@10 on \textit{ML10M}, while showing a decrease of 31.29\% in NDCG@10 and an increase of 59.58\% in PC@10 on \textit{CDs}. These results demonstrate that the proposed single diversity objective can improve diversity performance while maintaining accuracy performance, and outperforming other diversity methods like MMR and DPP.

Finally, compared with the single diversity objective optimized by CMB, the combined optimization objective also achieves a better balance between accuracy and diversity. For example, according to Table~\ref{rq11} and \ref{rq13}, on \textit{ML10M} and \textit{CDs}, the Recall@10 of $\text{CMB}_{\text{BPRMF}}$-SC-NDCG is 19.04\% and 35.71\% higher than $\text{CMB}_{\text{BPRMF}}$-SC, while the SC@10 only decreases by 5.06\% and 10.33\%, respectively. Similar findings can also be found in other cases. These results demonstrate the superiority of our proposed trade-off target in achieving a balance between accuracy and diversity.

\noindent \textbf{Diversification observations}.
First, the CMB model can effectively optimize various diversity metrics while yielding satisfactory results. For instance, as shown in Table~\ref{rq11} and \ref{rq13}, $\text{CMB}_{\text{BPRMF}}$-$\alpha$-nDCG outperforms the best baseline by 2.26\% in $\alpha$-nDCG@10 on \textit{ML10M}. Additionally, from these two tables, we also observe that $\text{CMB}_{\text{BPRMF}}$ significantly improves all diversity metrics compared to the base model BPRMF when optimized individually.

\begin{table*}
    \caption{Comparisons of the accuracy and diversity performance. The base model $g$ here adopts LightGCN. The bold scores are the best in each column, and the underlined scores are the second best. The symbols $\uparrow$ and $\downarrow$, along with their preceding values, represent the percentage (\% is omitted) improvement and decrease of a given method in the corresponding metric, in comparison to the base model $g$.} 
    \label{rq12}
    \centering
    \begin{adjustbox}{max width=\textwidth}
    \begin{tabular}{lllllll}
    \toprule
    Metric
    & Recall@10
    & NDCG@10
    & $\alpha$-nDCG@10
    & SC@10
    & PC@10
    & ILAD@10 \\
    \midrule
    \multicolumn{7}{c}{\textit{ML10M}} \\
    \midrule
        LightGCN & \textbf{0.1724} & \textbf{0.2912} & 0.7056 & \textbf{0.5680} & 0.1979 & 0.1480 \\
        \midrule
        $\text{CMB}_{\text{LightGCN}}$-$\alpha$-nDCG-NDCG & \underline{0.1706} (1.04$\downarrow$) & \underline{0.2866} (1.58$\downarrow$) & 0.7061 (0.07$\uparrow$) & 0.5655 (0.44$\downarrow$) & \underline{0.2131} (7.68$\uparrow$) & \underline{0.1562} (5.54$\uparrow$) \\
        $\text{CMB}_{\text{LightGCN}}$-SC-NDCG & \underline{0.1706} (1.04$\downarrow$) & 0.2865 (1.61$\downarrow$) & 0.7060 (0.06$\uparrow$) & 0.5650 (0.53$\downarrow$) & \textbf{0.2142} (8.24$\uparrow$) & \textbf{0.1565} (5.74$\uparrow$) \\
        $\text{CMB}_{\text{LightGCN}}$-PC-Recall & 0.1704 (1.16$\downarrow$) & \underline{0.2866} (1.58$\downarrow$) & \underline{0.7062} (0.09$\uparrow$) & \underline{0.5664} (0.28$\downarrow$) & 0.2122 (7.23$\uparrow$) & \underline{0.1562} (5.54$\uparrow$) \\
        $\text{CMB}_{\text{LightGCN}}$-ILAD-Recall & 0.1705 (1.10$\downarrow$) & \underline{0.2866} (1.58$\downarrow$) & \textbf{0.7072} (0.23$\uparrow$) & 0.5661 (0.33$\downarrow$) & 0.2130 (7.63$\uparrow$) & 0.1561 (5.47$\uparrow$) \\
    \midrule
    \multicolumn{7}{c}{\textit{CDs}} \\
    \midrule
        LightGCN & \textbf{0.0567} & \textbf{0.0500} & \textbf{0.7260} & 0.1616 & 0.0931 & 0.1659 \\
        \midrule
        $\text{CMB}_{\text{LightGCN}}$-$\alpha$-nDCG-Recall & \underline{0.0554} (2.29$\downarrow$) & \underline{0.0490} (2.00$\downarrow$) & \underline{0.7240} (0.28$\downarrow$) & \underline{0.1643} (1.67$\uparrow$) & \textbf{0.0938} (0.75$\uparrow$) & \textbf{0.1751} (5.55$\uparrow$) \\
        $\text{CMB}_{\text{LightGCN}}$-SC-Recall & \underline{0.0554} (2.29$\downarrow$) & 0.0489 (2.20$\downarrow$) & 0.7238 (0.30$\downarrow$) & 0.1642 (1.61$\uparrow$) & 0.0935 (0.43$\uparrow$) & \textbf{0.1751} (5.55$\uparrow$) \\
        $\text{CMB}_{\text{LightGCN}}$-PC-NDCG & 0.0553 (2.47$\downarrow$) & 0.0489 (2.20$\downarrow$) & 0.7238 (0.30$\downarrow$) & \textbf{0.1644} (1.73$\uparrow$) & 0.0934 (0.32$\uparrow$) & 0.1748 (5.36$\uparrow$) \\
        $\text{CMB}_{\text{LightGCN}}$-ILAD-NDCG & 0.0552 (2.65$\downarrow$) & 0.0488 (2.40$\downarrow$) & 0.7237 (0.32$\downarrow$) & 0.1642 (1.61$\uparrow$) & \underline{0.0936} (0.54$\uparrow$) & \underline{0.1750} (5.49$\uparrow$) \\
    \bottomrule
    \end{tabular}
    \end{adjustbox}
    \end{table*}
\begin{table*}
    \caption{Comparisons among the accuracy and diversity performance of CMB optimizes the single diversity metrics. The base model $g$ here adopts BPRMF. $\text{CMB}_{\text{ILAD}}^{\text{Gradient}}$ represents CMB that directly optimizes the differentiable metric ILAD by using the gradient method. $\text{CMB}_{\text{BPRMF}}$-Random represents CMB that chooses the arm randomly for each player. The bold scores are the best in each column.} 
    \label{rq13}
    \centering
    \begin{adjustbox}{max width=\textwidth}
    \begin{tabular}{lll|ll|ll|ll|ll|ll}
    \toprule
    \multirow{2}{*}{Metric}
    & \multicolumn{2}{c}{Recall@K}
    & \multicolumn{2}{c}{NDCG@K}
    & \multicolumn{2}{c}{$\alpha$-nDCG@K}
    & \multicolumn{2}{c}{SC@K}
    & \multicolumn{2}{c}{PC@K}
    & \multicolumn{2}{c}{ILAD@K} \\
    \cmidrule{2-3}
    \cmidrule{4-5}
    \cmidrule{6-7}
    \cmidrule{8-9}
    \cmidrule{10-11}
    \cmidrule{12-13}
    & \multicolumn{1}{c}{K=10}
    & \multicolumn{1}{c}{K=20}
    & \multicolumn{1}{c}{K=10}
    & \multicolumn{1}{c}{K=20}
    & \multicolumn{1}{c}{K=10}
    & \multicolumn{1}{c}{K=20}
    & \multicolumn{1}{c}{K=10}
    & \multicolumn{1}{c}{K=20}
    & \multicolumn{1}{c}{K=10}
    & \multicolumn{1}{c}{K=20}
    & \multicolumn{1}{c}{K=10}
    & \multicolumn{1}{c}{K=20} \\
    \midrule
    \multicolumn{13}{c}{\textit{ML10M}} \\
    \midrule
    $\text{CMB}_{\text{BPRMF}}$-Random & 0.1266 & 0.2032 & 0.2129 & 0.2217 & 0.7108 & 0.8045 & 0.5630 & 0.6930 & 0.2625 & 0.3514 & 0.2584 & 0.2787 \\
    $\text{CMB}_{\text{BPRMF}}$-$\alpha$-nDCG & \textbf{0.1267} & \textbf{0.2033} & \textbf{0.2138} & \textbf{0.2223} & \textbf{0.7467} & \textbf{0.8336} & 0.5794 & 0.7002 & 0.2620 & 0.3500 & 0.2554 & 0.2757 \\
    $\text{CMB}_{\text{BPRMF}}$-SC & 0.1250 & 0.2024 & 0.2106 & 0.2201 & 0.7436 & 0.8295 & \textbf{0.6065} & \textbf{0.7253} & 0.2607 & 0.3496 & 0.2542 & 0.2751 \\
    $\text{CMB}_{\text{BPRMF}}$-PC & 0.1259 & 0.2018 & 0.2135 & 0.2216 & 0.7104 & 0.8036 & 0.5726 & 0.7012 & 0.2714 & 0.3598 & 0.2582 & 0.2784 \\
    $\text{CMB}_{\text{BPRMF}}$-ILAD & 0.1258 & 0.2021 & 0.2121 & 0.2209 & 0.7147 & 0.8073 & 0.5650 & 0.6928 & 0.2651 & 0.3556 & 0.2652 & 0.2843 \\
    $\text{CMB}_{\text{BPRMF}}^{\text{Gradient}}$-ILAD & 0.1174 & 0.1914 & 0.1982 & 0.2086 & 0.6992 & 0.7948 & 0.5003 & 0.6279 &  \textbf{0.3089} & \textbf{0.3780} & \textbf{0.2900} & \textbf{0.3034} \\
    \midrule
    \multicolumn{13}{c}{\textit{CDs}} \\
    \midrule
    $\text{CMB}_{\text{BPRMF}}$-Random & 0.0353 & 0.0582 & 0.0314 & 0.0397 & 0.7053 & 0.8087 & 0.1850 & 0.2593 & 0.2659 & 0.3909 & 0.3081 & 0.3220 \\
    $\text{CMB}_{\text{BPRMF}}$-$\alpha$-nDCG & 0.0358 & 0.0583 & 0.0319 & 0.0400 & 0.7150 & \textbf{0.8155} & 0.1866 & 0.2593 & 0.2621 & 0.3859 & 0.3071 & 0.3209 \\
    $\text{CMB}_{\text{BPRMF}}$-SC & 0.0350 & 0.0579 & 0.0310 & 0.0393 & 0.7043 & 0.8074 & \textbf{0.1936} & \textbf{0.2684} & 0.2649 & 0.3884 & 0.3064 & 0.3205 \\
    $\text{CMB}_{\text{BPRMF}}$-PC & 0.0347 & 0.0572 & 0.0310 & 0.0391 & 0.7050 & 0.8086 & 0.1862 & 0.2610 & \textbf{0.2755} & \textbf{0.4050} & \textbf{0.3117} & \textbf{0.3255} \\
    $\text{CMB}_{\text{BPRMF}}$-ILAD & 0.0354 & 0.0585 & 0.0314 & 0.0397 & 0.7046 & 0.8083 & 0.1855 & 0.2598 & 0.2657 & 0.3885 & 0.3102 & 0.3233 \\
    $\text{CMB}_{\text{BPRMF}}^{\text{Gradient}}$-ILAD & \textbf{0.0524} & \textbf{0.0846} & \textbf{0.0458} & \textbf{0.0574} & \textbf{0.7164} & 0.8154 & 0.1717 & 0.2353 & 0.1768 & 0.2438 & 0.2788 & 0.2866 \\
    \bottomrule
    \end{tabular}
    \end{adjustbox}
    \end{table*}
Second, as illustrated in Tables~\ref{rq11} and \ref{rq13}, $\text{CMB}_{\text{BPRMF}}$-SC achieves the second-best SC@10 on \textit{ML10M}, only 0.12\% decrease compared to the best baseline. Similarly, $\text{CMB}_{\text{BPRMF}}$-PC also achieves the second-best performance in the PC metric on \textit{ML10M} and \textit{CDs}. These results demonstrate that CMB can achieve a satisfactory diversification. Moreover, while CMB sometimes trails DPP in diversity metrics, this is because DPP prioritizes diversity over accuracy. In contrast, our method balances both, leading to improved accuracy even if diversity scores are slightly lower when compared to the DPP.
\noindent \textbf{Application observations}.
First, from Table~\ref{rq13}, by comparing $\text{CMB}_{\text{BPRMF}}^{\text{Gradient}}$-ILAD and $\text{CMB}_{\text{BPRMF}}$-ILAD, we observe that the gradient descent optimization method outperforms the bandit optimization method in terms of accuracy on \textit{CDs}, while exhibiting better diversity results on \textit{ML10M} for PC and ILAD metrics. Thus, the choice of which optimization method should be based on the specific application scenario (e.g., dataset, diversity level, etc.).

Second, as shown in Table~\ref{rq11} and \ref{rq13}, compared to $\text{CMB}_{\text{Random}}$, CMB that only optimizes a single diversity metric also achieves good results on that metric, which shows that our bandit method is effective for optimizing different diversity metrics. Moreover, optimizing the combination of trade-off objectives simultaneously also achieves a good balance in accuracy and diversity, further highlighting the effectiveness of our combined optimization objectives. For example, NDCG@10 of $\text{CMB}_{\text{BPRMF}}$-SC-NDCG on \textit{ML10M} and \textit{CDs} is 18.88\% and 34.08\% higher than $\text{CMB}_{\text{BPRMF}}$-Random, respectively, while the SC@10 only decreases by 1.35\% and 6.16\%.

\noindent \textbf{Other observations.}
First, the accuracy and diversity trade-off exists widely. No method can achieve the best results in both accuracy and diversity since an increase in accuracy generally corresponds to a decrease in diversity. From the results in Table~\ref{rq11}, DPP achieves the best results in SC@10, PC@10, and ILAD@10 on \textit{ML1M} and \textit{ML10M}, but it achieves the worst performance in Recall@10 and NDCG@10 on these datasets.

Second, generally, no single method demonstrates superior performance across all diversity metrics. For example, as shown in Table~\ref{rq11}, just consider diversity metrics, DPP has the highest SC@10, PC@10, and ILAD@10 performance on \textit{ML10M} and \textit{CDs}, but has the lowest $\alpha$-nDCG@K performance. This indicates the inherent gap between different diversity evaluation metrics, proving the necessity of optimizing different metrics in a general framework, which is just the focus of our work.

    \begin{figure}[t]
	\centering
	\captionsetup[subfigure]{aboveskip=5pt,belowskip=5pt}
	\subcaptionbox{Comparisons performance of CMB when utilizing different erasure methods with top/least/random manners. \label{diff_erased}}
	{
		\includegraphics[width =0.49\linewidth]{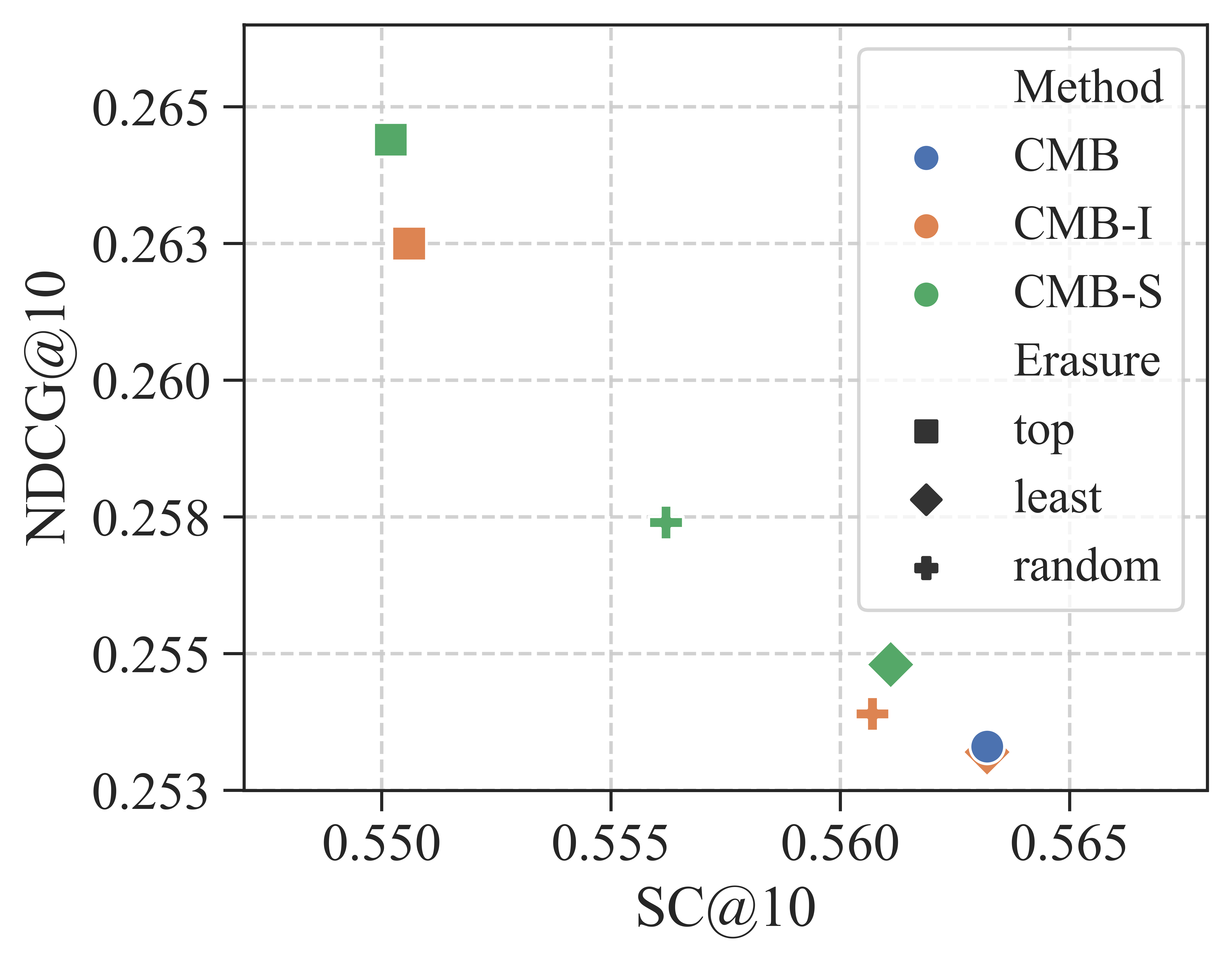}
		\includegraphics[width =0.49\linewidth]{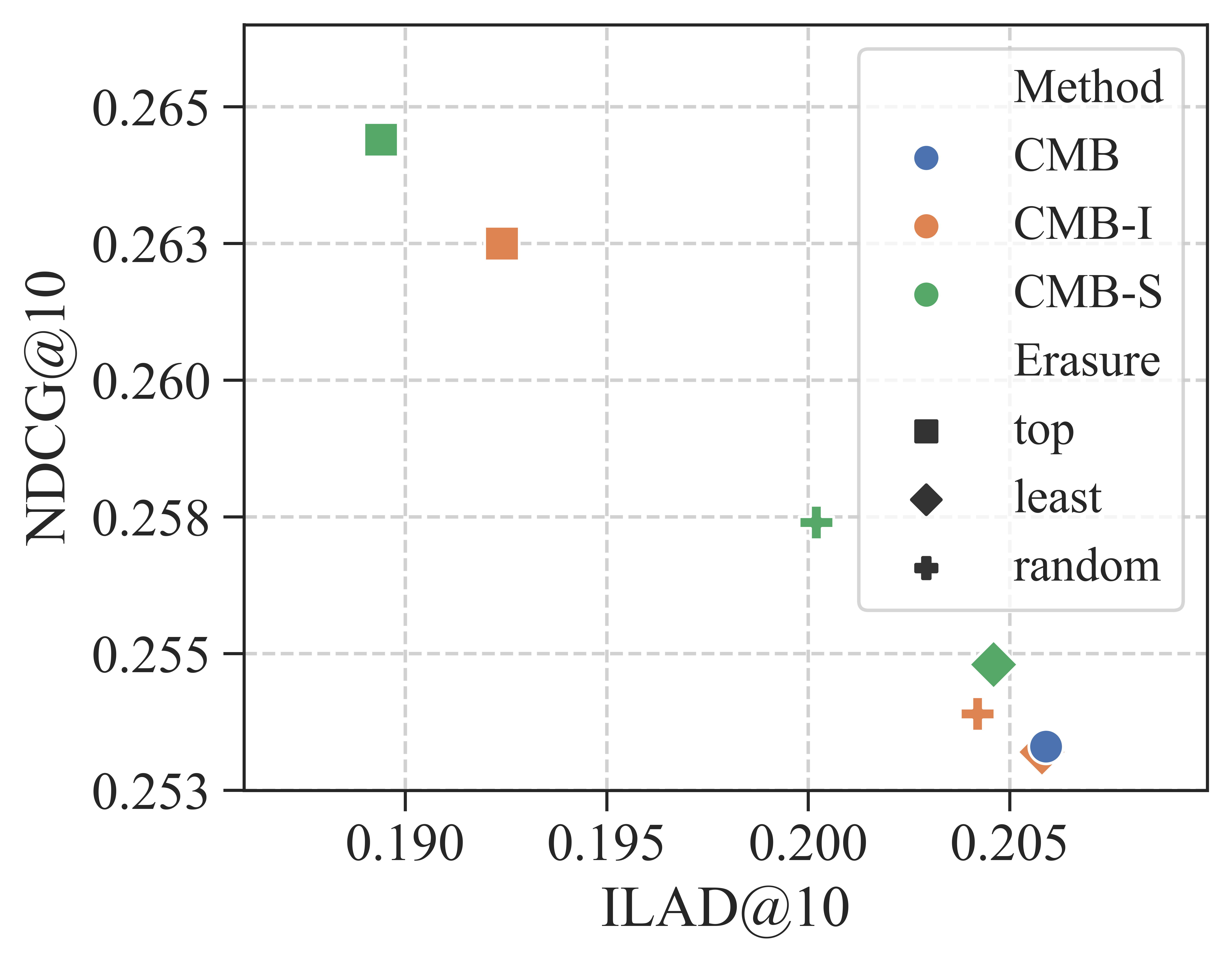}
	}
	\subcaptionbox{Comparisons performance of CMB when utilizing different erasure methods with different $F$. \label{diff-f}}
	{
		\includegraphics[width =0.49\linewidth]{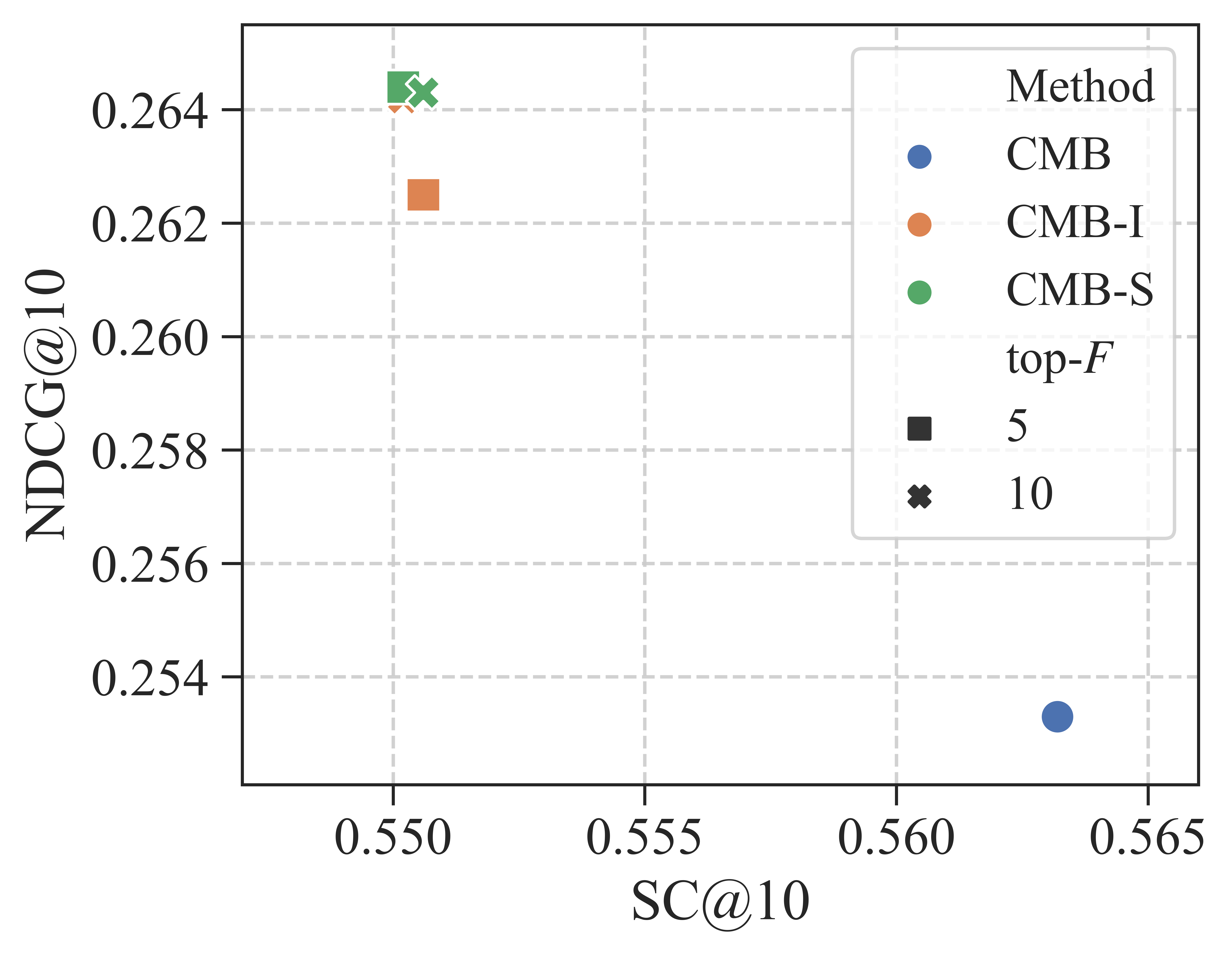}
		\includegraphics[width =0.49\linewidth]{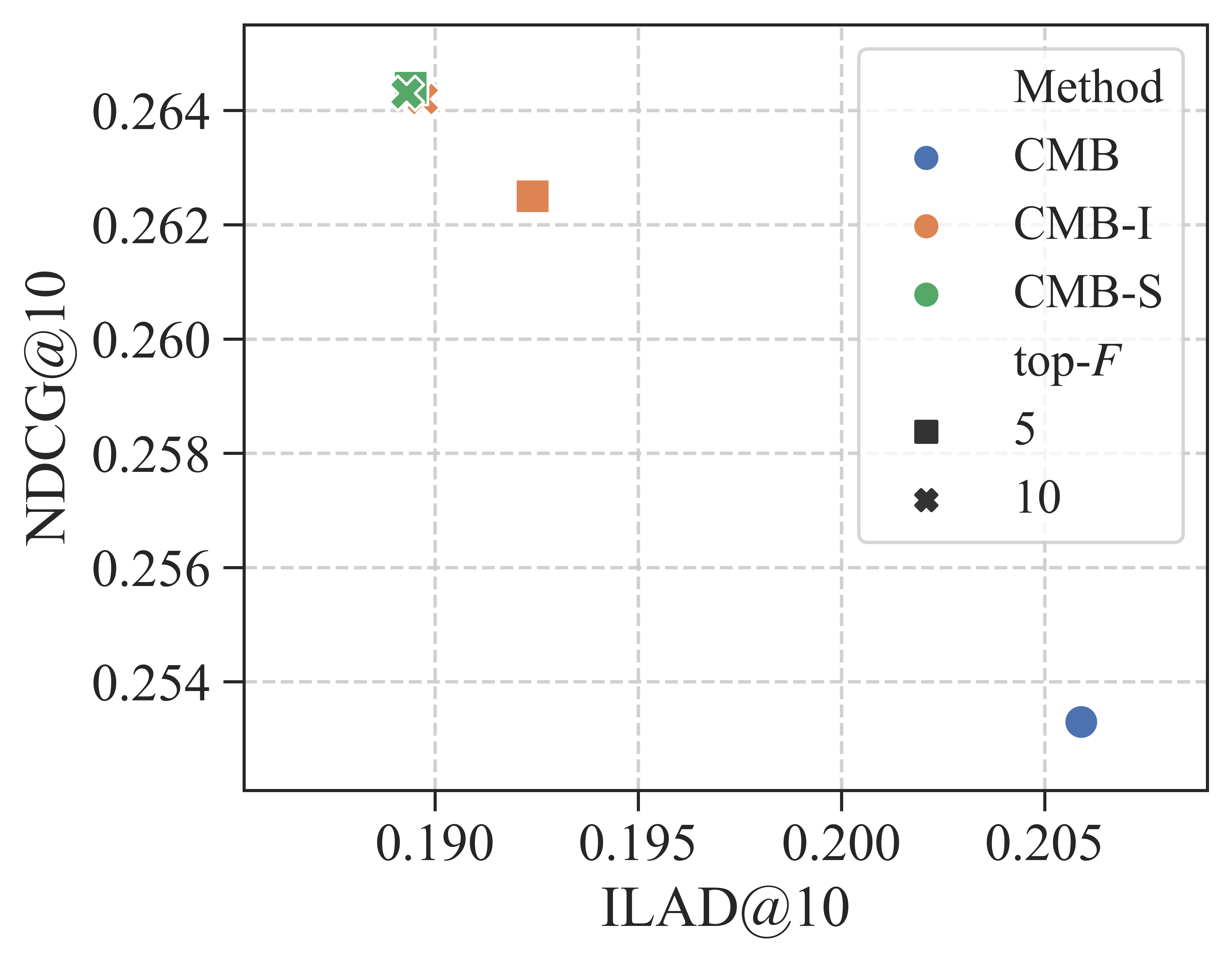}
	}
	\setlength{\abovecaptionskip}{5pt}
	\caption{Comparisons among the accuracy and diversity performance of CMB on dataset \textit{ML10M} when utilizing different erasure methods (CMB-I and CMB-S) with different erasure manners (top, least, and random) and different erasure features number $F$ (5 and 10).}
	\Description{Comparisons among the accuracy and diversity of CMB on \textit{ML10M} when utilizing different erasure methods with different manners and $F$.}
	\label{va}
\end{figure}

\subsection{Validity Analysis of Explanations (RQ2)}
As discussed in Sec.\ref{explainabtion}, the values of $\Delta$ affect the diversity of recommendation lists generated by the base model $g$. To evaluate whether $\Delta$ can discover the meaningful features that improve diversity or balance accuracy and diversity, we follow the widely deployed erasure-based evaluation criterion~\cite{DBLP:conf/emnlp/YuCZJ19,ge2022explainable} from Explainable AI. Specifically, we erase the ``most meaningful features'' from $\Delta$ (setting them to 0) and input this modified $\Delta$ into the pre-trained model $g$ to generate new recommendations. We then assess our model's effectiveness regarding the diversity and accuracy of these new results. We explore two erasure strategies, CMB-Individual (CMB-I) and CMB-Shared (CMB-S) -- by erasing the top, least, or random $F$ features, where $F$ is the number of erasing features. For CMB-I, we erase the top/least/random-$F$ features of each column of the absolute values of $\Delta$. For CMB-S, we average each row of the absolute values of $\Delta$, then erase the top/least/random-$F$ features by row. Compared with the least/random manners in Fig.~\ref{diff_erased}, we observe that omitting these meaningful features by the top manner reduces the diversity scores much while increasing the accuracy measures a lot. And the least manner does little to alter the performance of diversity or accuracy. Therefore, it verifies that the meaningful features we discover can benefit the trade-off between diversity and accuracy of recommendation results. Furthermore, as shown in Fig.~\ref{diff-f}, the results of the CMB-Shared approach with top-$F$ ($F$ = 5/10) are highly equivalent, indicating that the CMB-Shared approach can identify only a few features that significantly impact the model's diversity or accuracy.

\subsection{Case Study of Explanations (RQ3)}
\label{exp:case study}
\begin{figure}
	\centering
	\includegraphics[width=0.65\linewidth]{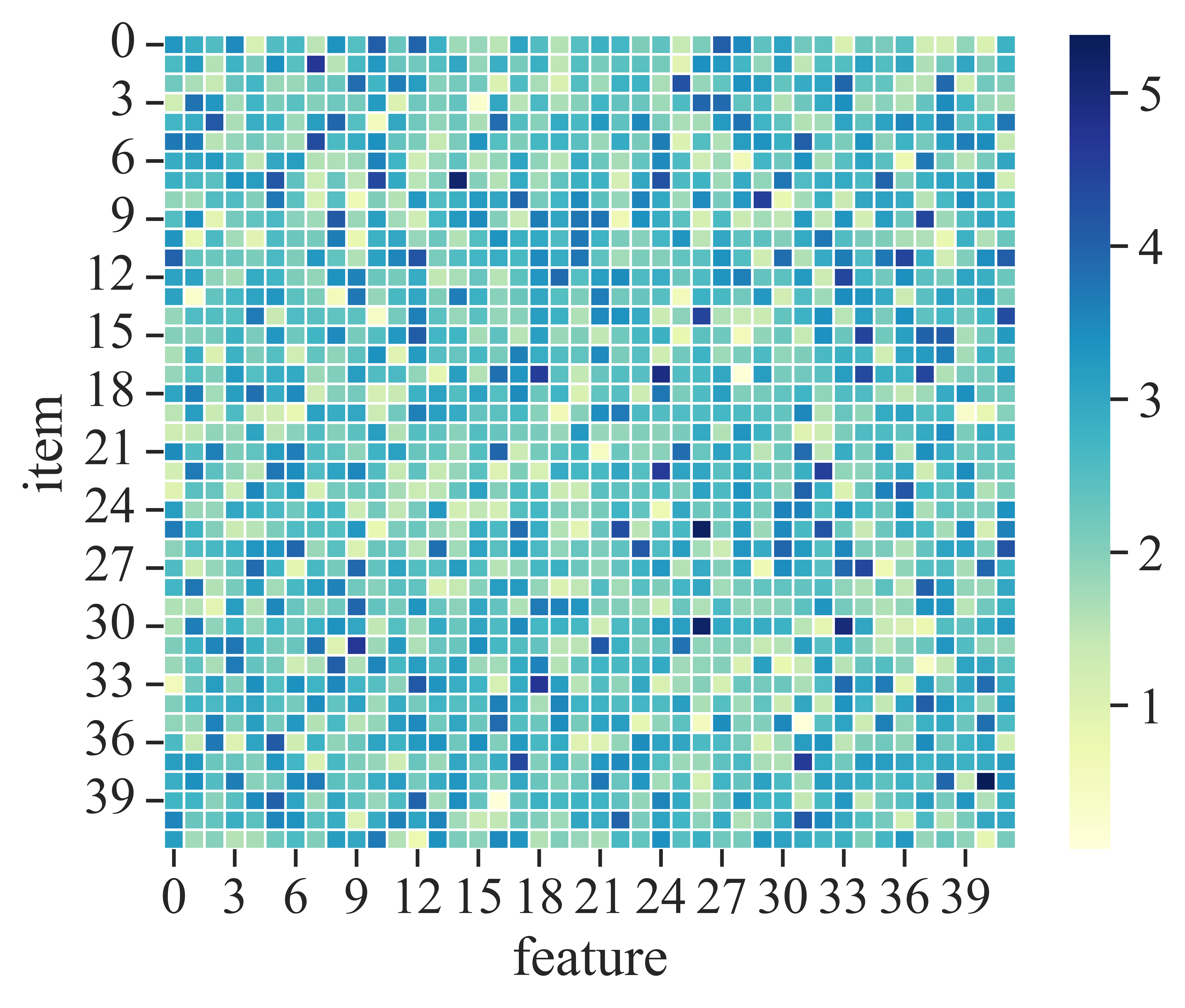}
	\caption{The feature explanations of CMB-Individual-$\alpha$-nDCG-Recall. Only the results of partial items are shown.}
	\Description{The feature explanations of CMB-Individual-$\alpha$-nDCG-Recall. Only the results of partial items are shown.}
	\label{cmb-Individual-map}
\end{figure}

The purpose of the case study is to demonstrate the
applicability of our method to both latent and raw features. As described in Sec.~\ref{explainabtion}, we illustrate how to generate explanations using raw features in this section. Following~\cite{ge2022explainable,zhang2014explicit}, we adopt the same method to extract the features and obtain the raw user and item feature matrices. Then, we apply two different feature-based explanations introduced in Sec.~\ref{explainabtion} on \textit{Phones} dataset from Amazon. The explanation results are presented in Fig.~\ref{cmb-Individual-map} and Table~\ref{tab:explanations}. These findings support our idea that it is challenging to manually discover feature explanations for diversity in recommender systems. For example, as shown in Table~\ref{tab:explanations}, it is difficult to know how input features (such as sound, charger, and connector) would determine the diversity of phone recommendations. As a result, explainable diversity approaches like ours are necessary to discover such features in the recommendation.

\begin{table}
	\centering
	\caption{Top-5 feature-based explanations on \textit{Phones} dataset. }
	\label{tab:explanations}
	\resizebox{\linewidth}{!}{
		\begin{tabular}{l l}
			\toprule
			{\bfseries Method} & {\bfseries Feature-based Explanations}\\
			\midrule
			CMB-Shared-SC-Recall & sound, volume, connector, headphone, protection \\
			CMB-Shared-SC-NDCG & charger, button, flashlight, cable, protection \\
			CMB-Shared-ILAD-Recall & connector, volume, pocket, charger, sound \\
			CMB-Shared-ILAD-NDCG & port, headset, plug, volume, package \\
			\bottomrule
		\end{tabular}
	}
\end{table}
\begin{figure}
	\centering
	\captionsetup[subfigure]{aboveskip=5pt,belowskip=5pt}
	\subcaptionbox{Comparisons performance of CMB when utilizing different $A$. \label{diff_a}}
	{
		\includegraphics[width =0.49\linewidth]{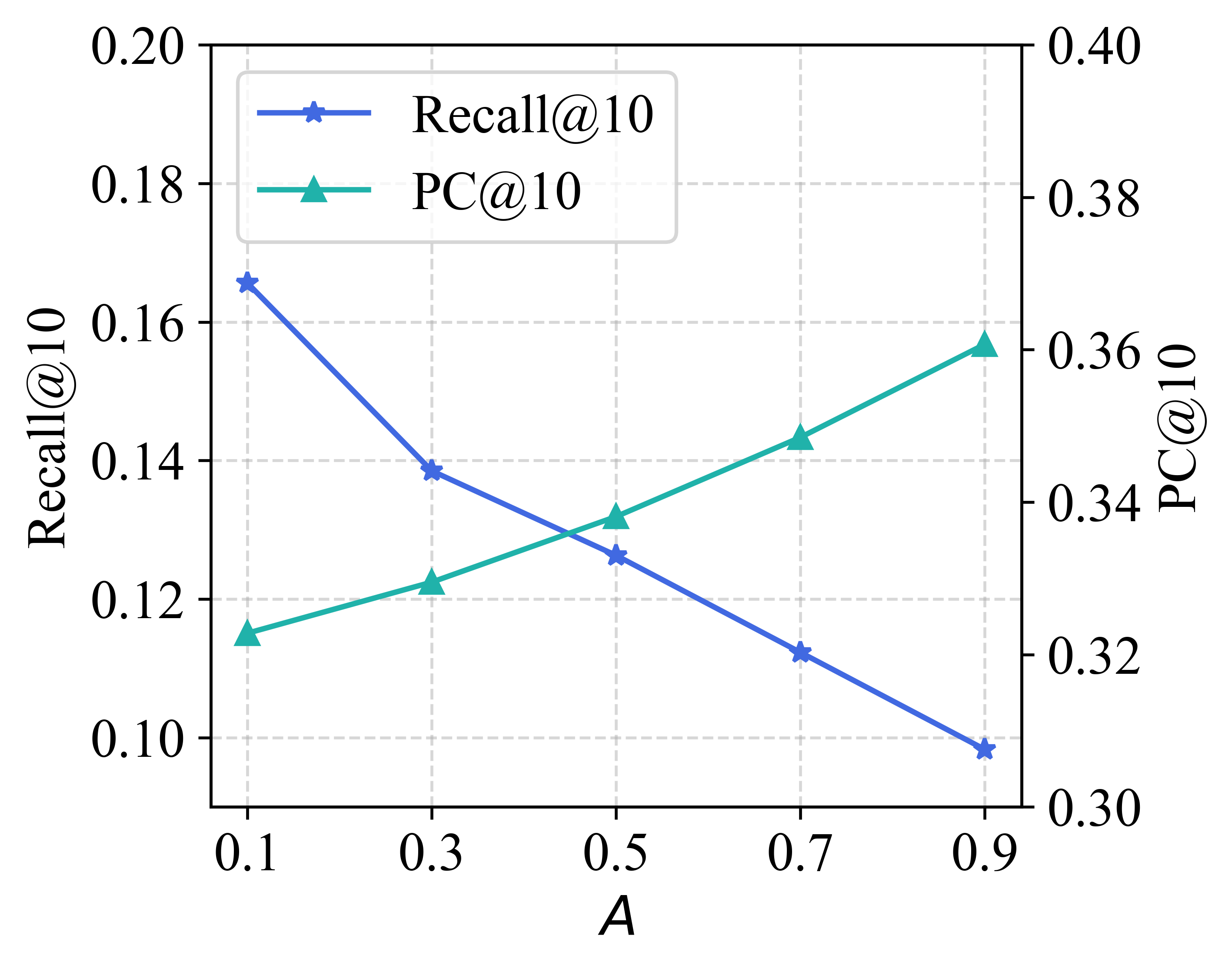}
		\includegraphics[width =0.49\linewidth]{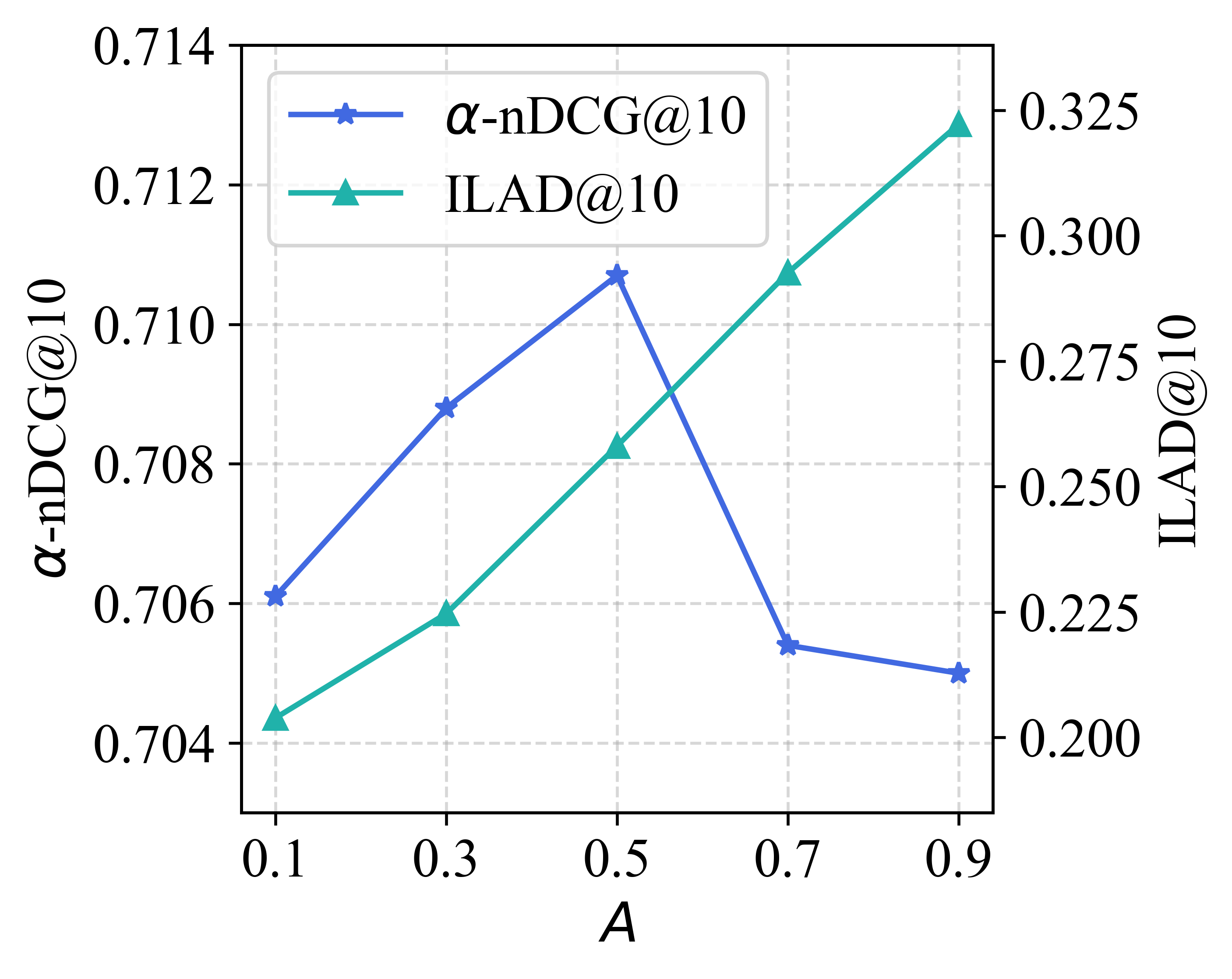}
	}
	\subcaptionbox{Comparisons performance of CMB when utilizing different $n_A$. \label{diff_na}}
	{
		\includegraphics[width =0.49\linewidth]{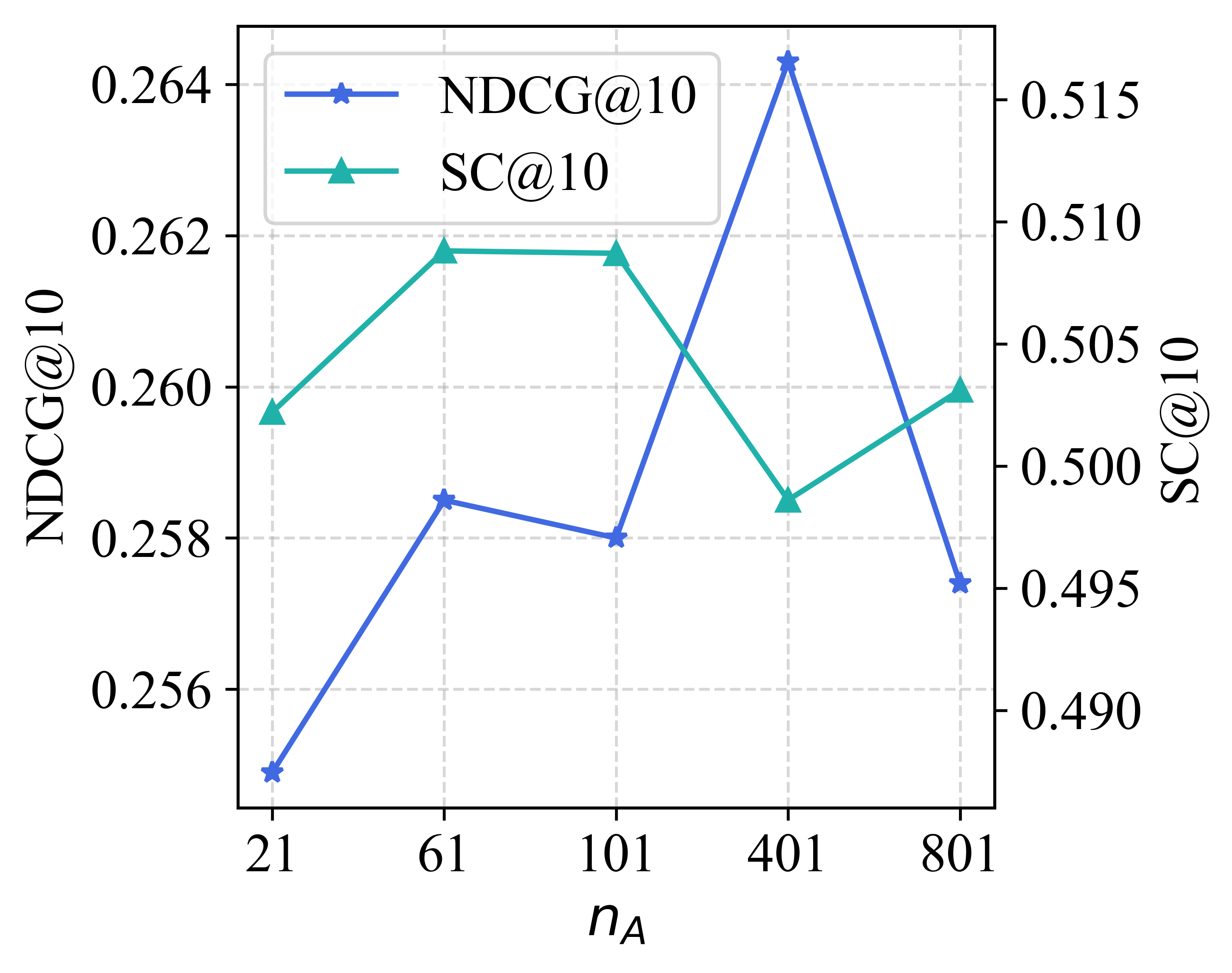}
		\includegraphics[width =0.49\linewidth]{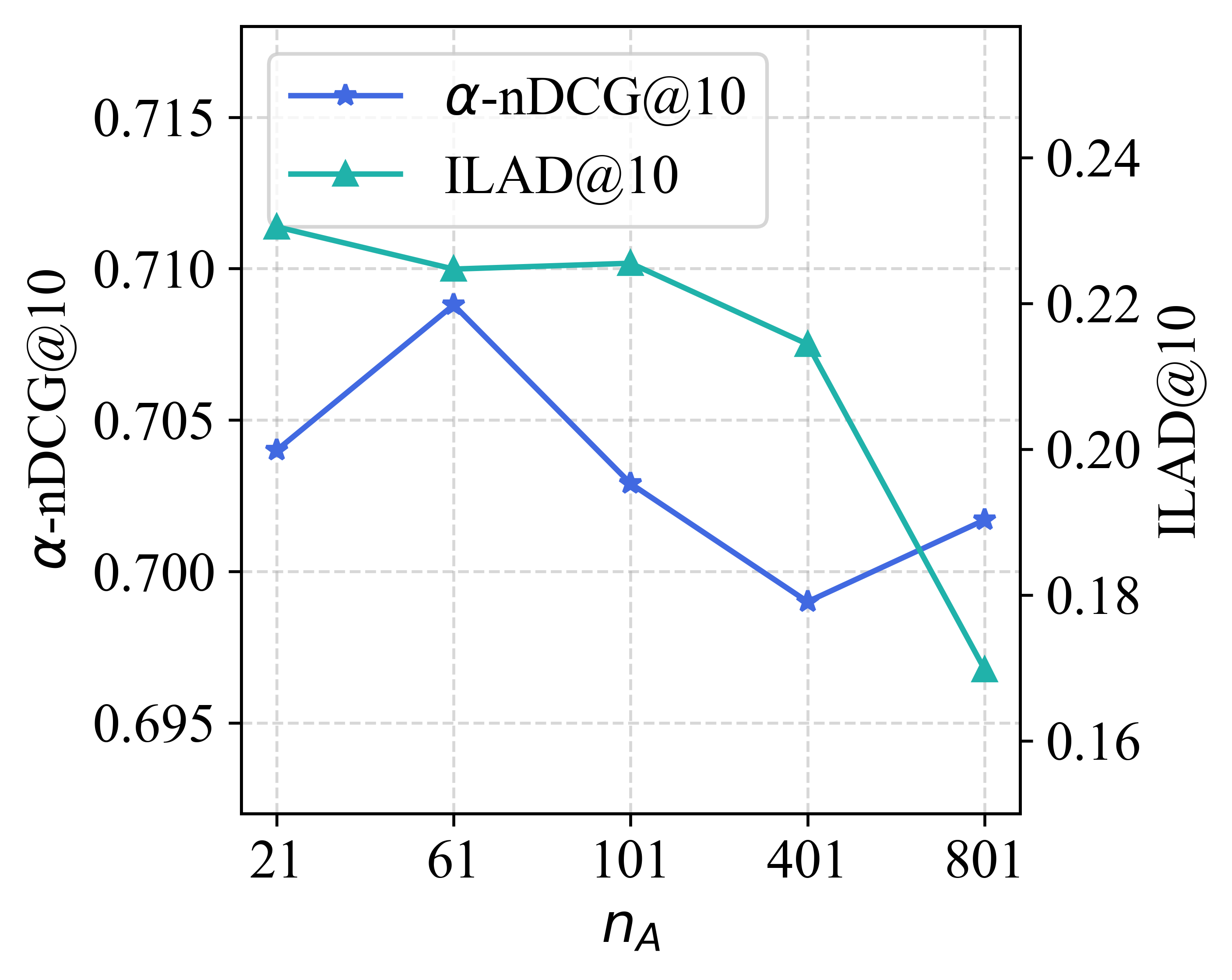}
	}
	\setlength{\abovecaptionskip}{5pt}
	\caption{Comparisons among the accuracy and diversity of CMB on dataset \textit{ML1M} when utilizing different $A$ and $n_A$.}
	\Description{Comparisons among the accuracy and diversity of CMB on dataset \textit{ML1M} when utilizing different $A$ and $n_A$.}
	\label{ab1}
\end{figure}

\subsection{The Sensitivity of Hyperparameters (RQ4)}
\label{armvalues}
The threshold of arm values $A$, and the number of arms $n_A$, are the two most important hyperparameters in the proposed model. Due to space limitations, we investigate their effects by $\text{CMB}_{\text{BPRMF}}$-ILAD-Recall on dataset \textit{ML1M} in Fig.~\ref{ab1}.
From the results in Fig.~\ref{diff_a}, we observe that when the number of arms $n_A$ is fixed ($n_A=61$), increasing $A$ results in a lower accuracy but a higher diversity performance. However, the impact on the $\alpha$-nDCG metric, which considers both accuracy and diversity, is more complex, as it first increases and then decreases with increasing $A$. The hyperparameter $A$ reflects the degree of the perturbations added, and should be chosen based on specific needs and requirements to determine an appropriate threshold.
Similarly, from the results in Fig.~\ref{diff_na}, when the threshold of arm values $A$ is fixed ($A=0.3$), the impact of the $n_A$ is more complex. For most metrics except ILAD, increasing $n_A$ leads to an initial increase in performance, followed by a decrease. However, the results of ILAD show a different trend.

\section{Conclusion and Future Work}
In this work, we propose CMB, a general bandit-based method, which optimizes the recommendation diversity while providing corresponding explanations. The method exhibits wide applicability and is agnostic to both recommendation models and diversity metrics. The proposed combination optimization target helps reach a more reasonable trade-off between recommendation accuracy and diversity performance. Besides, the explanations regarding diversification can be provided with the meaningfulness of the features obtained from counterfactual optimization. Extensive experiments on real-world datasets demonstrate our method's applicability, effectiveness, and explainability. In the future, we plan to design more efficient methods for generating explanations.

\begin{acks}
This work was supported by the Early Career Scheme (No. CityU 21219323) and the General Research Fund (No. CityU 11220324) of the University Grants Committee (UGC), and the NSFC Young Scientists Fund (No. 9240127).
\end{acks}

\bibliographystyle{ACM-Reference-Format.bst}
\bibliography{06_references.bib}


\appendix
\section{Experiments}
\subsection{Experiment Setup}
\subsubsection{\textbf{Datasets}}
\begin{itemize}[leftmargin=*]
    \item \textbf{MovieLens}~\cite{harper2015movielens}:
    This dataset contains user ratings on MovieLens web site. The MovieLens dataset includes several sub-datasets of varying sizes. For our evaluation, we select two datasets of different scales: \textit{MovieLens 1M} (\textit{ML1M}) and \textit{MovieLen 10M} (\textit{ML10M}), which offer the genre information we treat as the subtopic information.
    \item \textbf{Amazon}~\cite{DBLP:conf/emnlp/NiLM19}:
    This dataset contains user reviews on products in the Amazon e-commerce system. The Amazon dataset consists of 29 sub-datasets corresponding to different product categories. For our evaluation, we choose the dataset \textit{CDs and Vinyl} (\textit{CDs}).
\end{itemize}
\subsubsection{\textbf{Evaluation Metrics}}\leavevmode \\
\noindent \textbf{Accuracy}. We evaluate the accuracy of the ranking list using Recall@K and NDCG@K.
\begin{itemize}[leftmargin=*]
\item \textbf{Recall@K}: It indicates the coverage of the ground-truth (preferred) items as a result of top-$K$ recommendation. The value is computed by:
\begin{equation}
	\text{Recall@K}=\frac{|R_+^u \bigcap R^u|}{|R_+^u|},
\end{equation}
where the $R_+^u \in \mathcal{V}$ is an ordered set of the top-$K$ true (preferred) items, and $|R_+^u \bigcap R^u|$ is the number of true positives in recommendation list.

\item \textbf{NDCG@K}: The NDCG computes a score for $R^u$ which emphasizes higher-ranked true positives. The score is computed as:
\begin{equation}
\text{NDCG@K} = \frac{\text{DCG@K}}{\text{IDCG@K}}=\frac{\sum_{k=1}^{K}D(k)[v_l \in R_+^u]}{\sum_{k=1}^{K}D(k)},
\end{equation}
where $ D(k)=(2^{rel_k}-1)/log(k+1)$, and $rel_k$ is a relevance score at position $k$. We only consider binary responses, so we use a binary relevancy score: $rel_k=1$ if $v_k \in R_+^u$ and 0 otherwise.
\end{itemize}
\textbf{Diversity}. We evaluate the recommendation diversity by the $\alpha$-nDCG@K, SC@K, PC@K, and ILAD@K.
\begin{itemize}[leftmargin=*]
\item \textbf{Novelty-biased Normalized Discounted Cumulative Gain ($\alpha$-nDCG)}~\cite{clarke2008novelty}.
The $\alpha$-nDCG is a subtopic-level metric based on Normalized Discounted Cumulative Gain (NDCG) but considers the subtopic and is redundancy-aware along the items. The hyperparameter $\alpha$ is a geometric penalization for redundancy.
The formula for calculating $\alpha$-DCG@K is as follows:
\begin{equation}
    \text{$\alpha$-DCG@K} = \sum_{k=1}^{K}\frac{1}{\log(k+1)}\sum_{i=1}^{m}rel(v_k|s_i)(1-\alpha)^{cov(k, s_j)},
\end{equation}
where $cov(k, s_j)$ denotes the number of items covering subtopic 
$s_j$ till position $k$ in a ranking list $R^u$, $rel(v_k|s_i)$ is the relevance score of item $v_k$ for subtopic $s_i$ in position $k$, and $m$ is the total number of subtopics.
Then the $\alpha$-nDCG@K can be calculated by:
\begin{equation}
    \text{$\alpha$-nDCG@K} = \frac{\text{$\alpha$-DCG@K}}{\text{$\alpha$-IDCG@K}},
\end{equation}
where $\alpha$-IDCG@K = $max(\text{$\alpha$-DCG@K})$.

\item \textbf{Subtopic Coverage (SC)}~\cite{ge2010beyond, he2019diversity}. SC@K is a subtopic-level coverage of a recommended item list $R^u$ in the whole item set, which can be formulated as follows:
\begin{equation}
    \text{SC@K} = \frac{|\cup_{i \in R^u} \mathcal{S}(i)|}{\vert \cup_{i \in \mathcal{V}} \mathcal{S}(i)\vert},
\end{equation}
where $\mathcal{S}(i)$ is the set of subtopics covered by item $i$.

\item \textbf{Prediction Coverage (PC)}~\cite{ge2010beyond, herlocker2004evaluating}. PC@K is an item-level coverage of all recommendation lists $R^{u}$ in the whole item set, which can be formulated as follows:
     \begin{equation}
    \text{PC@K} = \frac{\vert \cup_{u \in \mathcal{U}} R^u \vert}{\vert \mathcal{V} \vert}.
    \end{equation}

\item \textbf{Intra-List Average Distance (ILAD)}~\cite{zhang2008avoiding}. ILAD@K is an item-level distance-based metric and evaluates the diversity by calculating the average dissimilarity of all pairs of items in the recommendation list $R^u$. In this work, we adopt cosine similarity to calculate the dissimilarity as the distance function.
\begin{equation}     \text{ILAD@K} = \underset{i, j \in R^u, i\neq j}{\text{mean}} (1 - cos\langle\mathbf{q}_i, \mathbf{q}_j\rangle).
\end{equation}
where $cos\langle\cdot, \cdot\rangle$ refers to the cosine similarity.
\end{itemize}

\subsection{Performance Comparison}

Tables~\ref{rq111}, \ref{rq121}, and \ref{rq131} are the full version results of the Tables~\ref{rq11}, \ref{rq12}, and \ref{rq13}, respectively.

    \begin{table*}
    \caption{Comparisons of the accuracy and diversity performance.
    The base model $g$ here adopts BPRMF.
    The bold scores are the best in each column, and the underlined scores are the second best. The symbols $\uparrow$ and $\downarrow$, along with their preceding values, represent the percentage (\% is omitted) improvement and decrease of a given method in the corresponding metric, in comparison to the base model $g$.} 
    \label{rq111}
    \centering
    \begin{adjustbox}{max width=\textwidth}
    \begin{tabular}{lrrrrrrrrrrrr}
    \toprule
    \multirow{2}{*}{Metrics}
    & \multicolumn{2}{c}{Recall@K}
    & \multicolumn{2}{c}{NDCG@K}
    & \multicolumn{2}{c}{$\alpha$-nDCG@K}
    & \multicolumn{2}{c}{SC@K}
    & \multicolumn{2}{c}{PC@K}
    & \multicolumn{2}{c}{ILAD@K} \\
    \cmidrule{2-3}
    \cmidrule{4-5}
    \cmidrule{6-7}
    \cmidrule{8-9}
    \cmidrule{10-11}
    \cmidrule{12-13}
    & \multicolumn{1}{c}{K=10}
    & \multicolumn{1}{c}{K=20}
    & \multicolumn{1}{c}{K=10}
    & \multicolumn{1}{c}{K=20}
    & \multicolumn{1}{c}{K=10}
    & \multicolumn{1}{c}{K=20}
    & \multicolumn{1}{c}{K=10}
    & \multicolumn{1}{c}{K=20}
    & \multicolumn{1}{c}{K=10}
    & \multicolumn{1}{c}{K=20}
    & \multicolumn{1}{c}{K=10}
    & \multicolumn{1}{c}{K=20} \\
    \midrule
    \multicolumn{13}{c}{\textit{ML1M}} \\
    \midrule
        BPRMF & \textbf{0.1465} & \textbf{0.2263} & \textbf{0.2742} & \textbf{0.2706} & 0.7035 & 0.7992 & 0.4993 & 0.6358 & 0.3206 & 0.4224 & 0.2010 & 0.2287  \\
        \midrule
        \multirow{2}{*}{MMR} & 0.0441 & 0.0790 & 0.0741 & 0.0787 & 0.6980 & 0.7947 & 0.4692 & 0.6006 & 0.0970 & 0.1619 & 0.1709 & 0.1983  \\
        ~ & 69.90$\downarrow$ & 65.09$\downarrow$ & 72.98$\downarrow$ & 70.92$\downarrow$ & 0.78$\downarrow$ & 0.56$\downarrow$ & 6.03$\downarrow$ & 5.54$\downarrow$ & 69.74$\downarrow$ & 61.67$\downarrow$ & 14.98$\downarrow$ & 13.29$\downarrow$  \\
        \hline
        \multirow{2}{*}{DPP} & 0.0360 & 0.0613 & 0.0689 & 0.0723 & \textbf{0.7186} & \textbf{0.8188} & \textbf{0.5558} & \textbf{0.7060} & \textbf{0.4554} & \textbf{0.5184} & \textbf{0.3693} & \textbf{0.3648}  \\
        ~ & 75.43$\downarrow$ & 72.91$\downarrow$ & 74.87$\downarrow$ & 73.28$\downarrow$ & 2.15$\uparrow$ & 2.45$\uparrow$ & 11.32$\uparrow$ & 11.04$\uparrow$ & 42.05$\uparrow$ & 22.73$\uparrow$ & 83.73$\uparrow$ & 59.51$\uparrow$  \\
        \hline
        \multirow{2}{*}{$\text{CMB}_{\text{BPRMF}}$-$\alpha$-nDCG-Recall} & 0.1388 & 0.2161 & 0.2588 & 0.2567 & 0.7094 & 0.8034 & 0.5097 & 0.6440 & 0.3291 & 0.4307 & \underline{0.2253} & \underline{0.2511}  \\
        ~ & 5.26$\downarrow$ & 4.51$\downarrow$ & 5.62$\downarrow$ & 5.14$\downarrow$ & 0.84$\uparrow$ & 0.53$\uparrow$ & 2.08$\uparrow$ & 1.29$\uparrow$ & 2.65$\uparrow$ & 1.96$\uparrow$ & 12.09$\uparrow$ & 9.79$\uparrow$  \\
        \hline
        \multirow{2}{*}{$\text{CMB}_{\text{BPRMF}}$-SC-Recall} & \underline{0.1391} & \underline{0.2171} & \underline{0.2594} & \underline{0.2575} & 0.7078 & 0.8019 & 0.5102 & 0.6448 & 0.3307 & 0.4321 & 0.2246 & 0.2506  \\
        ~ & 5.05$\downarrow$ & 4.07$\downarrow$ & 5.40$\downarrow$ & 4.84$\downarrow$ & 0.61$\uparrow$ & 0.34$\uparrow$ & 2.18$\uparrow$ & 1.42$\uparrow$ & 3.15$\uparrow$ & 2.30$\uparrow$ & 11.74$\uparrow$ & 9.58$\uparrow$  \\
        \hline
        \multirow{2}{*}{$\text{CMB}_{\text{BPRMF}}$-PC-NDCG} & 0.1388 & 0.2162 & 0.2584 & 0.2564 & 0.7093 & 0.8032 & 0.5107 & 0.6451 & 0.3286 & \underline{0.4330} & 0.2247 & 0.2506  \\
        ~ & 5.26$\downarrow$ & 4.46$\downarrow$ & 5.76$\downarrow$ & 5.25$\downarrow$ & 0.82$\uparrow$ & 0.50$\uparrow$ & 2.28$\uparrow$ & 1.46$\uparrow$ & 2.50$\uparrow$ & 2.51$\uparrow$ & 11.79$\uparrow$ & 9.58$\uparrow$  \\
        \hline
        \multirow{2}{*}{$\text{CMB}_{\text{BPRMF}}$-ILAD-NDCG} & 0.1387 & 0.2165 & 0.2587 & 0.2567 & \underline{0.7109} & \underline{0.8050} & \underline{0.5127} & \underline{0.6479} & \underline{0.3313} & 0.4312 & 0.2246 & 0.2504  \\
        ~ & 5.32$\downarrow$ & 4.33$\downarrow$ & 5.65$\downarrow$ & 5.14$\downarrow$ & 1.05$\uparrow$ & 0.73$\uparrow$ & 2.68$\uparrow$ & 1.90$\uparrow$ & 3.34$\uparrow$ & 2.08$\uparrow$ & 11.74$\uparrow$ & 9.49$\uparrow$ \\
        \midrule
    \multicolumn{13}{c}{\textit{ML10M}} \\
    \midrule
        BPRMF & \textbf{0.1549} & \textbf{0.2416} & \textbf{0.2648} & \textbf{0.2709} & 0.7043 & 0.7998 & 0.5483 & 0.6801 & 0.2453 & 0.3236 & 0.1886 & 0.2128 \\
        \midrule
        \multirow{2}{*}{MMR} & 0.0402 & 0.0663 & 0.0602 & 0.0658 & 0.7095 & \textbf{0.8076} & 0.5155 & 0.6500 & 0.0416 & 0.0697 & 0.1623 & 0.1867  \\
        ~ & 74.05$\downarrow$ & 72.56$\downarrow$ & 77.27$\downarrow$ & 75.71$\downarrow$ & 0.74$\uparrow$ & 0.98$\uparrow$ & 5.98$\downarrow$ & 4.43$\downarrow$ & 83.04$\downarrow$ & 78.46$\downarrow$ & 13.94$\downarrow$ & 12.27$\downarrow$  \\
        \hline
        \multirow{2}{*}{DPP} & 0.0253 & 0.0398 & 0.0541 & 0.0537 & 0.6977 & 0.7952 & \textbf{0.6072} & \textbf{0.7451} & \textbf{0.3735} & \textbf{0.4085} & \textbf{0.3764} & \textbf{0.3645}  \\
        ~ & 83.67$\downarrow$ & 83.53$\downarrow$ & 79.57$\downarrow$ & 80.18$\downarrow$ & 0.94$\downarrow$ & 0.58$\downarrow$ & 10.74$\uparrow$ & 9.56$\uparrow$ & 52.26$\uparrow$ & 26.24$\uparrow$ & 99.58$\uparrow$ & 71.29$\uparrow$  \\
        \hline
        \multirow{2}{*}{$\text{CMB}_{\text{BPRMF}}$-$\alpha$-nDCG-Recall} & 0.1488 & 0.2337 & 0.2531 & 0.2600 & \textbf{0.7126} & \underline{0.8066} & 0.5544 & 0.6843 & 0.2475 & 0.3291 & 0.2045 & 0.2275  \\
        ~ & 3.94$\downarrow$ & 3.27$\downarrow$ & 4.42$\downarrow$ & 4.02$\downarrow$ & 1.18$\uparrow$ & 0.85$\uparrow$ & 1.11$\uparrow$ & 0.62$\uparrow$ & 0.90$\uparrow$ & 1.70$\uparrow$ & 8.43$\uparrow$ & 6.91$\uparrow$  \\
        \hline
        \multirow{2}{*}{$\text{CMB}_{\text{BPRMF}}$-SC-Recall} & \underline{0.1490} & \underline{0.2339} & \underline{0.2535} & \underline{0.2604} & \underline{0.7123} & 0.8063 & 0.5544 & 0.6843 & \underline{0.2477} & \underline{0.3304} & 0.2046 & 0.2276  \\
        ~ & 3.81$\downarrow$ & 3.19$\downarrow$ & 4.27$\downarrow$ & 3.88$\downarrow$ & 1.14$\uparrow$ & 0.81$\uparrow$ & 1.11$\uparrow$ & 0.62$\uparrow$ & 0.98$\uparrow$ & 2.10$\uparrow$ & 8.48$\uparrow$ & 6.95$\uparrow$  \\
        \hline
        \multirow{2}{*}{$\text{CMB}_{\text{BPRMF}}$-PC-NDCG} & 0.1487 & 0.2336 & 0.2532 & 0.2600 & \underline{0.7123} & 0.8063 & \underline{0.5546} & 0.6843 & 0.2471 & 0.3292 & \underline{0.2051} & \underline{0.2280}  \\
        ~ & 4.00$\downarrow$ & 3.31$\downarrow$ & 4.38$\downarrow$ & 4.02$\downarrow$ & 1.14$\uparrow$ & 0.81$\uparrow$ & 1.15$\uparrow$ & 0.62$\uparrow$ & 0.73$\uparrow$ & 1.73$\uparrow$ & 8.75$\uparrow$ & 7.14$\uparrow$  \\
        \hline
        \multirow{2}{*}{$\text{CMB}_{\text{BPRMF}}$-ILAD-NDCG} & 0.1486 & 0.2336 & 0.2527 & 0.2598 & 0.7110 & 0.8052 & 0.5541 & \underline{0.6844} & 0.2460 & 0.3300 & 0.2050 & 0.2279  \\
        ~ & 4.07$\downarrow$ & 3.31$\downarrow$ & 4.57$\downarrow$ & 4.10$\downarrow$ & 0.95$\uparrow$ & 0.68$\uparrow$ & 1.06$\uparrow$ & 0.63$\uparrow$ & 0.29$\uparrow$ & 1.98$\uparrow$ & 8.70$\uparrow$ & 7.10$\uparrow$ \\
    \midrule
    \multicolumn{13}{c}{\textit{CDs}} \\
    \midrule
        BPRMF & \textbf{0.0515} & \textbf{0.0838} & \textbf{0.0457} & \textbf{0.0575} & \underline{0.7206} & \underline{0.8174} & 0.1700 & 0.2312 & 0.1665 & 0.2342 & 0.2332 & 0.2466  \\
        \midrule
        \multirow{2}{*}{MMR} & 0.0033 & 0.0114 & 0.0032 & 0.0061 & \textbf{0.7240} & \textbf{0.8311} & 0.1705 & 0.2335 & 0.0247 & 0.0424 & 0.2372 & 0.2520  \\
        ~ & 93.59$\downarrow$ & 86.40$\downarrow$ & 93.00$\downarrow$ & 89.39$\downarrow$ & 0.47$\uparrow$ & 1.68$\uparrow$ & 0.29$\uparrow$ & 0.99$\uparrow$ & 85.17$\downarrow$ & 81.90$\downarrow$ & 1.72$\uparrow$ & 2.19$\uparrow$  \\
        \hline
        \multirow{2}{*}{DPP} & 0.0115 & 0.0170 & 0.0128 & 0.0146 & 0.7116 & 0.8165 & \textbf{0.2409} & \textbf{0.3440} & \textbf{0.3261} & \textbf{0.4353} & \textbf{0.4013} & \textbf{0.3966}  \\
        ~ & 77.67$\downarrow$ & 79.71$\downarrow$ & 71.99$\downarrow$ & 74.61$\downarrow$ & 1.25$\downarrow$ & 0.11$\downarrow$ & 41.71$\uparrow$ & 48.79$\uparrow$ & 95.86$\uparrow$ & 85.87$\uparrow$ & 72.08$\uparrow$ & 60.83$\uparrow$  \\
        \hline
        \multirow{2}{*}{$\text{CMB}_{\text{BPRMF}}$-$\alpha$-nDCG-NDCG} & \underline{0.0477} & 0.0776 & \underline{0.0422} & \underline{0.0531} & 0.7183 & 0.8163 & \underline{0.1739} & \underline{0.2374} & \underline{0.1825} & 0.2625 & 0.2511 & 0.2646  \\
        ~ & 7.38$\downarrow$ & 7.40$\downarrow$ & 7.66$\downarrow$ & 7.65$\downarrow$ & 0.32$\downarrow$ & 0.13$\downarrow$ & 2.29$\uparrow$ & 2.68$\uparrow$ & 9.61$\uparrow$ & 12.08$\uparrow$ & 7.68$\uparrow$ & 7.30$\uparrow$  \\
        \hline
        \multirow{2}{*}{$\text{CMB}_{\text{BPRMF}}$-SC-NDCG} & 0.0475 & 0.0777 & 0.0421 & \underline{0.0531} & 0.7192 & 0.8169 & 0.1736 & 0.2368 & 0.1824 & 0.2620 & 0.2510 & 0.2645  \\
        ~ & 7.77$\downarrow$ & 7.28$\downarrow$ & 7.88$\downarrow$ & 7.65$\downarrow$ & 0.19$\downarrow$ & 0.06$\downarrow$ & 2.12$\uparrow$ & 2.42$\uparrow$ & 9.55$\uparrow$ & 11.87$\uparrow$ & 7.63$\uparrow$ & 7.26$\uparrow$  \\
       \hline
        \multirow{2}{*}{$\text{CMB}_{\text{BPRMF}}$-PC-Recall} & 0.0476 & 0.0778 & 0.0421 & \underline{0.0531} & 0.7180 & 0.8162 & 0.1737 & 0.2371 & 0.1816 & 0.2623 & 0.2509 & 0.2644  \\
        ~ & 7.57$\downarrow$ & 7.16$\downarrow$ & 7.88$\downarrow$ & 7.65$\downarrow$ & 0.36$\downarrow$ & 0.15$\downarrow$ & 2.18$\uparrow$ & 2.55$\uparrow$ & 9.07$\uparrow$ & 12.00$\uparrow$ & 7.59$\uparrow$ & 7.22$\uparrow$  \\
        \hline
        \multirow{2}{*}{$\text{CMB}_{\text{BPRMF}}$-ILAD-Recall} & \underline{0.0477} & \underline{0.0779} & \underline{0.0422} & \underline{0.0531} & 0.7189 & 0.8167 & 0.1736 & 0.2369 & 0.1823 & \underline{0.2627} & \underline{0.2513} & \underline{0.2649}  \\
        ~ & 7.38$\downarrow$ & 7.04$\downarrow$ & 7.66$\downarrow$ & 7.65$\downarrow$ & 0.24$\downarrow$ & 0.09$\downarrow$ & 2.12$\uparrow$ & 2.47$\uparrow$ & 9.49$\uparrow$ & 12.17$\uparrow$ & 7.76$\uparrow$ & 7.42$\uparrow$ \\
    \bottomrule
    \end{tabular}
    \end{adjustbox}
    \end{table*}

    \begin{table*}
    \caption{Comparisons of the accuracy and diversity performance.
    The base model $g$ here adopts LightGCN.
    The bold scores are the best in each column, and the underlined scores are the second best. The symbols $\uparrow$ and $\downarrow$, along with their preceding values, represent the percentage (\% is omitted) improvement and decrease of a given method in the corresponding metric, in comparison to the base model $g$.} 
    \label{rq121}
    \centering
    \begin{adjustbox}{max width=\textwidth}
    \begin{tabular}{lrrrrrrrrrrrr}
    \toprule
    \multirow{2}{*}{Metrics}
    & \multicolumn{2}{c}{Recall@K}
    & \multicolumn{2}{c}{NDCG@K}
    & \multicolumn{2}{c}{$\alpha$-nDCG@K}
    & \multicolumn{2}{c}{SC@K}
    & \multicolumn{2}{c}{PC@K}
    & \multicolumn{2}{c}{ILAD@K} \\
    \cmidrule{2-3}
    \cmidrule{4-5}
    \cmidrule{6-7}
    \cmidrule{8-9}
    \cmidrule{10-11}
    \cmidrule{12-13}
    & \multicolumn{1}{c}{K=10}
    & \multicolumn{1}{c}{K=20}
    & \multicolumn{1}{c}{K=10}
    & \multicolumn{1}{c}{K=20}
    & \multicolumn{1}{c}{K=10}
    & \multicolumn{1}{c}{K=20}
    & \multicolumn{1}{c}{K=10}
    & \multicolumn{1}{c}{K=20}
    & \multicolumn{1}{c}{K=10}
    & \multicolumn{1}{c}{K=20}
    & \multicolumn{1}{c}{K=10}
    & \multicolumn{1}{c}{K=20} \\
    \midrule
    \multicolumn{13}{c}{\textit{ML1M}} \\
    \midrule
        LightGCN & \textbf{0.1623} & \textbf{0.2478} & \textbf{0.2884} & \textbf{0.2878} & 0.6844 & 0.7861 & 0.4655 & 0.6033 & \textbf{0.4038} & \textbf{0.5190} & 0.1832 & 0.2024  \\
        \midrule
        $\text{CMB}_{\text{LightGCN}}$-$\alpha$-nDCG-NDCG & \underline{0.1583} & 0.2420 & \underline{0.2822} & 0.2812 & \textbf{0.6889} & \textbf{0.7894} & \textbf{0.4780} & \textbf{0.6159} & 0.3933 & 0.5055 & 0.1860 & 0.2047  \\
        ~ & 2.46$\downarrow$ & 2.34$\downarrow$ & 2.15$\downarrow$ & 2.29$\downarrow$ & 0.66$\uparrow$ & 0.42$\uparrow$ & 2.69$\uparrow$ & 2.09$\uparrow$ & 2.60$\downarrow$ & 2.60$\downarrow$ & 1.53$\uparrow$ & 1.14$\uparrow$  \\
        \hline
        $\text{CMB}_{\text{LightGCN}}$-SC-NDCG & 0.1576 & \underline{0.2421} & 0.2814 & 0.2810 & 0.6883 & 0.7889 & \underline{0.4765} & \underline{0.6147} & \underline{0.3934} & \underline{0.5082} & \textbf{0.1878} & \textbf{0.2066}  \\
        ~ & 2.90$\downarrow$ & 2.30$\downarrow$ & 2.43$\downarrow$ & 2.36$\downarrow$ & 0.57$\uparrow$ & 0.36$\uparrow$ & 2.36$\uparrow$ & 1.89$\uparrow$ & 2.58$\downarrow$ & 2.08$\downarrow$ & 2.51$\uparrow$ & 2.08$\uparrow$  \\
        \hline
        $\text{CMB}_{\text{LightGCN}}$-PC-Recall & 0.1578 & \underline{0.2421} & 0.2820 & \underline{0.2813} & \underline{0.6885} & \underline{0.7891} & 0.4762 & 0.6141 & 0.3932 & \underline{0.5082} & \underline{0.1877} & \underline{0.2063}  \\
        ~ & 2.77$\downarrow$ & 2.30$\downarrow$ & 2.22$\downarrow$ & 2.26$\downarrow$ & 0.60$\uparrow$ & 0.38$\uparrow$ & 2.30$\uparrow$ & 1.79$\uparrow$ & 2.63$\downarrow$ & 2.08$\downarrow$ & 2.46$\uparrow$ & 1.93$\uparrow$  \\
        \hline
        $\text{CMB}_{\text{LightGCN}}$-ILAD-Recall & \underline{0.1583} & 0.2418 & \underline{0.2822} & 0.2811 & 0.6883 & 0.7890 & \underline{0.4765} & 0.6145 & 0.3933 & 0.5065 & 0.1870 & 0.2057  \\
        ~ & 2.46$\downarrow$ & 2.42$\downarrow$ & 2.15$\downarrow$ & 2.33$\downarrow$ & 0.57$\uparrow$ & 0.37$\uparrow$ & 2.36$\uparrow$ & 1.86$\uparrow$ & 2.60$\downarrow$ & 2.41$\downarrow$ & 2.07$\uparrow$ & 1.63$\uparrow$ \\
    \midrule
    \multicolumn{13}{c}{\textit{ML10M}} \\
    \midrule
        LightGCN & \textbf{0.1724} & \textbf{0.2659} & \textbf{0.2912} & \textbf{0.2978} & 0.7056 & 0.8006 & \textbf{0.5680} & \textbf{0.6995} & 0.1979 & 0.2643 & 0.1480 & 0.1677  \\
        \midrule
        \multirow{2}{*}{$\text{CMB}_{\text{LightGCN}}$-$\alpha$-nDCG-NDCG} & \underline{0.1706} & 0.2634 & \underline{0.2866} & 0.2934 & 0.7061 & 0.8009 & 0.5655 & 0.6969 & \underline{0.2131} & \textbf{0.2832} & \underline{0.1562} & \underline{0.1762}  \\
        ~ & 1.04$\downarrow$ & 0.94$\downarrow$ & 1.58$\downarrow$ & 1.48$\downarrow$ & 0.07$\uparrow$ & 0.04$\uparrow$ & 0.44$\downarrow$ & 0.37$\downarrow$ & 7.68$\uparrow$ & 7.15$\uparrow$ & 5.54$\uparrow$ & 5.07$\uparrow$  \\
        \hline
        \multirow{2}{*}{$\text{CMB}_{\text{LightGCN}}$-SC-NDCG} & \underline{0.1706} & \underline{0.2635} & 0.2865 & \underline{0.2935} & 0.7060 & 0.8009 & 0.5650 & 0.6965 & \textbf{0.2142} & \underline{0.2825} & \textbf{0.1565} & \textbf{0.1764}  \\
        ~ & 1.04$\downarrow$ & 0.90$\downarrow$ & 1.61$\downarrow$ & 1.44$\downarrow$ & 0.06$\uparrow$ & 0.04$\uparrow$ & 0.53$\downarrow$ & 0.43$\downarrow$ & 8.24$\uparrow$ & 6.89$\uparrow$ & 5.74$\uparrow$ & 5.19$\uparrow$  \\
        \hline
        \multirow{2}{*}{$\text{CMB}_{\text{LightGCN}}$-PC-Recall} & 0.1704 & 0.2633 & \underline{0.2866} & \underline{0.2935} & \underline{0.7062} & \underline{0.8010} & \underline{0.5664} & \underline{0.6976} & 0.2122 & 0.2817 & \underline{0.1562} & 0.1761  \\
        ~ & 1.16$\downarrow$ & 0.98$\downarrow$ & 1.58$\downarrow$ & 1.44$\downarrow$ & 0.09$\uparrow$ & 0.05$\uparrow$ & 0.28$\downarrow$ & 0.27$\downarrow$ & 7.23$\uparrow$ & 6.58$\uparrow$ & 5.54$\uparrow$ & 5.01$\uparrow$  \\
        \hline
        \multirow{2}{*}{$\text{CMB}_{\text{LightGCN}}$-ILAD-Recall} & 0.1705 & \underline{0.2635} & \underline{0.2866} & \underline{0.2935} & \textbf{0.7072} & \textbf{0.8019} & 0.5661 & 0.6971 & 0.2130 & 0.2822 & 0.1561 & 0.1761  \\
        ~ & 1.10$\downarrow$ & 0.90$\downarrow$ & 1.58$\downarrow$ & 1.44$\downarrow$ & 0.23$\uparrow$ & 0.16$\uparrow$ & 0.33$\downarrow$ & 0.34$\downarrow$ & 7.63$\uparrow$ & 6.77$\uparrow$ & 5.47$\uparrow$ & 5.01$\uparrow$ \\
    \midrule
    \multicolumn{13}{c}{\textit{CDs}} \\
    \midrule
        LightGCN & \textbf{0.0567} & \textbf{0.0927} & \textbf{0.0500} & \textbf{0.0630} & \textbf{0.7260} & \textbf{0.8210} & 0.1616 & 0.2188 & 0.0931 & 0.1425 & 0.1659 & 0.1802  \\
        \midrule
        \multirow{2}{*}{$\text{CMB}_{\text{LightGCN}}$-$\alpha$-nDCG-Recall} & \underline{0.0554} & \underline{0.0901} & \underline{0.0490} & \underline{0.0614} & \underline{0.7240} & \underline{0.8197} & \underline{0.1643} & \textbf{0.2230} & \textbf{0.0938} & \textbf{0.1450} & \textbf{0.1751} & \textbf{0.1898}  \\
        ~ & 2.29$\downarrow$ & 2.80$\downarrow$ & 2.00$\downarrow$ & 2.54$\downarrow$ & 0.28$\downarrow$ & 0.16$\downarrow$ & 1.67$\uparrow$ & 1.92$\uparrow$ & 0.75$\uparrow$ & 1.75$\uparrow$ & 5.55$\uparrow$ & 5.33$\uparrow$  \\
        \hline
        \multirow{2}{*}{$\text{CMB}_{\text{LightGCN}}$-SC-Recall} & \underline{0.0554} & 0.0900 & 0.0489 & \underline{0.0614} & 0.7238 & 0.8196 & 0.1642 & \underline{0.2229} & 0.0935 & \textbf{0.1450} & \textbf{0.1751} & \textbf{0.1898}  \\
        ~ & 2.29$\downarrow$ & 2.91$\downarrow$ & 2.20$\downarrow$ & 2.54$\downarrow$ & 0.30$\downarrow$ & 0.17$\downarrow$ & 1.61$\uparrow$ & 1.87$\uparrow$ & 0.43$\uparrow$ & 1.75$\uparrow$ & 5.55$\uparrow$ & 5.33$\uparrow$  \\
        \hline
        \multirow{2}{*}{$\text{CMB}_{\text{LightGCN}}$-PC-NDCG} & 0.0553 & 0.0899 & 0.0489 & 0.0613 & 0.7238 & 0.8195 & \textbf{0.1644} & \textbf{0.2230} & 0.0934 & 0.1445 & 0.1748 & 0.1895  \\
        ~ & 2.47$\downarrow$ & 3.02$\downarrow$ & 2.20$\downarrow$ & 2.70$\downarrow$ & 0.30$\downarrow$ & 0.18$\downarrow$ & 1.73$\uparrow$ & 1.92$\uparrow$ & 0.32$\uparrow$ & 1.40$\uparrow$ & 5.36$\uparrow$ & 5.16$\uparrow$  \\
        \hline
        \multirow{2}{*}{$\text{CMB}_{\text{LightGCN}}$-ILAD-NDCG} & 0.0552 & \underline{0.0901} & 0.0488 & \underline{0.0614} & 0.7237 & 0.8196 & 0.1642 & \underline{0.2229} & \underline{0.0936} & \underline{0.1447} & \underline{0.1750} & \underline{0.1896}  \\
        ~ & 2.65$\downarrow$ & 2.80$\downarrow$ & 2.40$\downarrow$ & 2.54$\downarrow$ & 0.32$\downarrow$ & 0.17$\downarrow$ & 1.61$\uparrow$ & 1.87$\uparrow$ & 0.54$\uparrow$ & 1.54$\uparrow$ & 5.49$\uparrow$ & 5.22$\uparrow$ \\
    \bottomrule
    \end{tabular}
    \end{adjustbox}
    \end{table*}

\begin{table*}
    \caption{Comparisons among the accuracy and diversity performance of CMB optimizes the single diversity metrics.
    The base model $g$ here adopts BPRMF.
    $\text{CMB}_{\text{ILAD}}^{\text{Gradient}}$ represents CMB that directly optimizes the differentiable metric ILAD by using the gradient method. $\text{CMB}_{\text{BPRMF}}$-Random represents CMB that chooses the arm randomly for each player. The bold scores are the best in each column.} 
    \label{rq131}
    \centering
    \begin{adjustbox}{max width=\textwidth}
    \begin{tabular}{lll|ll|ll|ll|ll|ll}
    \toprule
    \multirow{2}{*}{Metric}
    & \multicolumn{2}{c}{Recall@K}
    & \multicolumn{2}{c}{NDCG@K}
    & \multicolumn{2}{c}{$\alpha$-nDCG@K}
    & \multicolumn{2}{c}{SC@K}
    & \multicolumn{2}{c}{PC@K}
    & \multicolumn{2}{c}{ILAD@K} \\
    \cmidrule{2-3}
    \cmidrule{4-5}
    \cmidrule{6-7}
    \cmidrule{8-9}
    \cmidrule{10-11}
    \cmidrule{12-13}
    & \multicolumn{1}{c}{K=10}
    & \multicolumn{1}{c}{K=20}
    & \multicolumn{1}{c}{K=10}
    & \multicolumn{1}{c}{K=20}
    & \multicolumn{1}{c}{K=10}
    & \multicolumn{1}{c}{K=20}
    & \multicolumn{1}{c}{K=10}
    & \multicolumn{1}{c}{K=20}
    & \multicolumn{1}{c}{K=10}
    & \multicolumn{1}{c}{K=20}
    & \multicolumn{1}{c}{K=10}
    & \multicolumn{1}{c}{K=20} \\
    \midrule
    \multicolumn{13}{c}{\textit{ML1M}} \\
    \midrule
    $\text{CMB}_{\text{BPRMF}}$-Random & \textbf{0.1099} & 0.1769 & \textbf{0.2055} & \textbf{0.2068} & 0.7001 & 0.7976 & 0.5192 & 0.6597 & 0.3568 & 0.4688 & 0.2992 & 0.3200 \\
    $\text{CMB}_{\text{BPRMF}}$-$\alpha$-nDCG & 0.1098 & 0.1770 & 0.2050 & 0.2062 & \textbf{0.7403} & \textbf{0.8275} & 0.5436 & 0.6733 & 0.3572 & 0.4717 & 0.2941 & 0.3158 \\
    $\text{CMB}_{\text{BPRMF}}$-SC & 0.1097 & \textbf{0.1774} & 0.2048 & 0.2061 & 0.7253 & 0.8157 & \textbf{0.5725} & \textbf{0.7035} & 0.3541 & 0.4729 & 0.2941 & 0.3163 \\
    $\text{CMB}_{\text{BPRMF}}$-PC & 0.1085 & 0.1756 & 0.2030 & 0.2049 & 0.6955 & 0.7928 & 0.5185 & 0.6575 & 0.3765 & 0.4862 & 0.3022 & 0.3227 \\
    $\text{CMB}_{\text{BPRMF}}$-ILAD & 0.1072 & 0.1732 & 0.2000 & 0.2024 & 0.6938 & 0.7925 & 0.5026 & 0.6423 & 0.3638 & 0.4762 & 0.3079 & 0.3266 \\
    $\text{CMB}_{\text{BPRMF}}^{\text{Gradient}}$-ILAD & 0.0972 & 0.1631 & 0.1778 & 0.1854 & 0.6677 & 0.7741 & 0.4324 & 0.5695 & \textbf{0.5057} & \textbf{0.5830} & \textbf{0.3552} & \textbf{0.3682} \\
    \midrule
    \multicolumn{13}{c}{\textit{ML10M}} \\
    \midrule
    $\text{CMB}_{\text{BPRMF}}$-Random & 0.1266 & 0.2032 & 0.2129 & 0.2217 & 0.7108 & 0.8045 & 0.5630 & 0.6930 & 0.2625 & 0.3514 & 0.2584 & 0.2787 \\
    $\text{CMB}_{\text{BPRMF}}$-$\alpha$-nDCG & \textbf{0.1267} & \textbf{0.2033} & \textbf{0.2138} & \textbf{0.2223} & \textbf{0.7467} & \textbf{0.8336} & 0.5794 & 0.7002 & 0.2620 & 0.3500 & 0.2554 & 0.2757 \\
    $\text{CMB}_{\text{BPRMF}}$-SC & 0.1250 & 0.2024 & 0.2106 & 0.2201 & 0.7436 & 0.8295 & \textbf{0.6065} & \textbf{0.7253} & 0.2607 & 0.3496 & 0.2542 & 0.2751 \\
    $\text{CMB}_{\text{BPRMF}}$-PC & 0.1259 & 0.2018 & 0.2135 & 0.2216 & 0.7104 & 0.8036 & 0.5726 & 0.7012 & 0.2714 & 0.3598 & 0.2582 & 0.2784 \\
    $\text{CMB}_{\text{BPRMF}}$-ILAD & 0.1258 & 0.2021 & 0.2121 & 0.2209 & 0.7147 & 0.8073 & 0.5650 & 0.6928 & 0.2651 & 0.3556 & 0.2652 & 0.2843 \\
    $\text{CMB}_{\text{BPRMF}}^{\text{Gradient}}$-ILAD & 0.1174 & 0.1914 & 0.1982 & 0.2086 & 0.6992 & 0.7948 & 0.5003 & 0.6279 &  \textbf{0.3089} & \textbf{0.3780} & \textbf{0.2900} & \textbf{0.3034} \\
    \midrule
    \multicolumn{13}{c}{\textit{CDs}} \\
    \midrule
    $\text{CMB}_{\text{BPRMF}}$-Random & 0.0353 & 0.0582 & 0.0314 & 0.0397 & 0.7053 & 0.8087 & 0.1850 & 0.2593 & 0.2659 & 0.3909 & 0.3081 & 0.3220 \\
    $\text{CMB}_{\text{BPRMF}}$-$\alpha$-nDCG & 0.0358 & 0.0583 & 0.0319 & 0.0400 & 0.7150 & \textbf{0.8155} & 0.1866 & 0.2593 & 0.2621 & 0.3859 & 0.3071 & 0.3209 \\
    $\text{CMB}_{\text{BPRMF}}$-SC & 0.0350 & 0.0579 & 0.0310 & 0.0393 & 0.7043 & 0.8074 & \textbf{0.1936} & \textbf{0.2684} & 0.2649 & 0.3884 & 0.3064 & 0.3205 \\
    $\text{CMB}_{\text{BPRMF}}$-PC & 0.0347 & 0.0572 & 0.0310 & 0.0391 & 0.7050 & 0.8086 & 0.1862 & 0.2610 & \textbf{0.2755} & \textbf{0.4050} & \textbf{0.3117} & \textbf{0.3255} \\
    $\text{CMB}_{\text{BPRMF}}$-ILAD & 0.0354 & 0.0585 & 0.0314 & 0.0397 & 0.7046 & 0.8083 & 0.1855 & 0.2598 & 0.2657 & 0.3885 & 0.3102 & 0.3233 \\
    $\text{CMB}_{\text{BPRMF}}^{\text{Gradient}}$-ILAD & \textbf{0.0524} & \textbf{0.0846} & \textbf{0.0458} & \textbf{0.0574} & \textbf{0.7164} & 0.8154 & 0.1717 & 0.2353 & 0.1768 & 0.2438 & 0.2788 & 0.2866 \\
    \bottomrule
    \end{tabular}
    \end{adjustbox}
    \end{table*}

\subsection{The Sensitivity of Hyperparameters}

Table~\ref{hy1} and \ref{hy2} are the full version results of the sensitivity of hyperparameters $A$ and $n_A$, respectively. Table~\ref{hy3} shows the results of the sensitivity of hyperparameters $\lambda_2$ by $\text{CMB}_{\text{BPRMF}}$-ILAD-Recall on dataset \textit{ML1M}. Based on the results presented in Table~\ref{hy3}, we observe that as the value of $\lambda_2$ increases, the accuracy performance generally improves while the diversity performance decreases. However, the scale of these changes is relatively small, indicating that the proposed combination trade-off objective is highly robust and that the influence of the $\lambda_2$ parameter is limited.

\begin{table*}
    \caption{Comparisons among the accuracy and diversity of CMB on dataset \textit{ML1M} when utilizing different $A$.}
    \label{hy1}
    \centering
    \begin{adjustbox}{max width=\textwidth}
    \begin{tabular}{lcc|cc|cc|cc|cc|cc}
    \toprule
    \multirow{2}{*}{Metrics}
    & \multicolumn{2}{c}{Recall@K}
    & \multicolumn{2}{c}{NDCG@K}
    & \multicolumn{2}{c}{$\alpha$-nDCG@K}
    & \multicolumn{2}{c}{SC@K}
    & \multicolumn{2}{c}{PC@K}
    & \multicolumn{2}{c}{ILAD@K} \\
    \cmidrule{2-3}
    \cmidrule{4-5}
    \cmidrule{6-7}
    \cmidrule{8-9}
    \cmidrule{10-11}
    \cmidrule{12-13}
    & \multicolumn{1}{c}{K=5}
    & \multicolumn{1}{c}{K=10}
    & \multicolumn{1}{c}{K=5}
    & \multicolumn{1}{c}{K=10}
    & \multicolumn{1}{c}{K=5}
    & \multicolumn{1}{c}{K=10}
    & \multicolumn{1}{c}{K=5}
    & \multicolumn{1}{c}{K=10}
    & \multicolumn{1}{c}{K=5}
    & \multicolumn{1}{c}{K=10}
    & \multicolumn{1}{c}{K=5}
    & \multicolumn{1}{c}{K=10}
    \\
    \midrule
    $A = 0.1, n_A = 61$ & \textbf{0.0897} & \textbf{0.1459} & \textbf{0.2926} & \textbf{0.2726} & 0.6460 & 0.7061 & 0.3744 & 0.5017 & 0.2398 & 0.3228 & 0.1816 & 0.2039 \\
    $A = 0.3, n_A = 61$ & 0.0842 & 0.1385 & 0.2760 & 0.2585 & 0.6491 & 0.7088 & 0.3806 & 0.5088 & 0.2424 & 0.3295 & 0.2038 & 0.2247 \\
    $A = 0.5, n_A = 61$ & 0.0758 & 0.1263 & 0.2492 & 0.2346 & \textbf{0.6504} & \textbf{0.7107} & 0.3892 & 0.5203 & 0.2498 & 0.3381 & 0.2391 & 0.2581 \\
    $A = 0.7, n_A = 61$ & 0.0673 & 0.1123 & 0.2200 & 0.2079 & 0.6441 & 0.7054 & 0.3902 & 0.5255 & 0.2572 & 0.3485 & 0.2763 & 0.2926 \\
    $A = 0.9, n_A = 61$ & 0.0577 & 0.0983 & 0.1905 & 0.1811 & 0.6438 & 0.7050 & \textbf{0.3965} & \textbf{0.5351} & \textbf{0.2632} & \textbf{0.3607} & \textbf{0.3076} & \textbf{0.3222} \\
    \bottomrule
    \end{tabular}
    \end{adjustbox}
    \end{table*}

\begin{table*}
    \caption{Comparisons among the accuracy and diversity of CMB on dataset \textit{ML1M} when utilizing different $n_A$.}
    \label{hy2}
    \centering
    \begin{adjustbox}{max width=\textwidth}
    \begin{tabular}{lcc|cc|cc|cc|cc|cc}
    \toprule
    \multirow{2}{*}{Metrics}
    & \multicolumn{2}{c}{Recall@K}
    & \multicolumn{2}{c}{NDCG@K}
    & \multicolumn{2}{c}{$\alpha$-nDCG@K}
    & \multicolumn{2}{c}{SC@K}
    & \multicolumn{2}{c}{PC@K}
    & \multicolumn{2}{c}{ILAD@K} \\
    \cmidrule{2-13}
    & \multicolumn{1}{c}{K=5}
    & \multicolumn{1}{c}{K=10}
    & \multicolumn{1}{c}{K=5}
    & \multicolumn{1}{c}{K=10}
    & \multicolumn{1}{c}{K=5}
    & \multicolumn{1}{c}{K=10}
    & \multicolumn{1}{c}{K=5}
    & \multicolumn{1}{c}{K=10}
    & \multicolumn{1}{c}{K=5}
    & \multicolumn{1}{c}{K=10}
    & \multicolumn{1}{c}{K=5}
    & \multicolumn{1}{c}{K=10}
    \\
    \midrule
    $A = 0.3, n_A = 21$ & 0.0833 & 0.1367 & 0.2729 & 0.2549 & 0.6427 & 0.7040 & 0.3734 & 0.5022 & \textbf{0.2500} & \textbf{0.3318} & \textbf{0.2103} & \textbf{0.2305} \\
    $A = 0.3, n_A = 41$ & 0.0838 & 0.1379 & 0.2753 & 0.2576 & 0.6434 & 0.7047 & 0.3750 & 0.5056 & 0.2445 & 0.3247 & 0.2057 & 0.2257 \\
    $A = 0.3, n_A = 61$ & 0.0842 & 0.1385 & 0.2760 & 0.2585 & \textbf{0.6491} & \textbf{0.7088} & \textbf{0.3806} & 0.5088 & 0.2424 & 0.3295 & 0.2038 & 0.2247 \\
    $A = 0.3, n_A = 81$ & 0.0843 & 0.1382 & 0.2769 & 0.2584 & 0.6421 & 0.7038 & 0.3787 & \textbf{0.5096} & 0.2425 & 0.3261 & 0.2039 & 0.2249 \\
    $A = 0.3, n_A = 101$ & 0.0842 & 0.1383 & 0.2763 & 0.2580 & 0.6409 & 0.7029 & 0.3763 & 0.5087 & 0.2445 & 0.3267 & 0.2049 & 0.2255 \\
    $A = 0.3, n_A = 201$ & 0.0791 & 0.1317 & 0.2582 & 0.2434 & 0.6338 & 0.6968 & 0.3734 & 0.5086 & 0.2481 & 0.3311 & 0.1295 & 0.1373 \\
    $A = 0.3, n_A = 401$ & \textbf{0.0871} & \textbf{0.1420} & \textbf{0.2832} & \textbf{0.2643} & 0.6360 & 0.6990 & 0.3689 & 0.4986 & 0.2417 & 0.3211 & 0.1938 & 0.2144 \\
    $A = 0.3, n_A = 601$ & 0.0850 & 0.1391 & 0.2778 & 0.2602 & 0.6418 & 0.7027 & 0.3712 & 0.5004 & 0.2453 & 0.3270 & 0.1723 & 0.1888 \\
    $A = 0.3, n_A = 801$ & 0.0837 & 0.1378 & 0.2752 & 0.2574& 0.6403 & 0.7017 & 0.3726 & 0.5031 & 0.2454 & 0.3300 & 0.1561 & 0.1698 \\
    $A = 0.3, n_A = 1001$ & 0.0833 & 0.1365 & 0.2721 & 0.2546 & 0.6369 & 0.6994 & 0.3707 & 0.5008 & 0.2435 & 0.3248 & 0.1496 & 0.1613 \\
    \bottomrule
    \end{tabular}
    \end{adjustbox}
    \end{table*}

    \begin{table*}
    \caption{Comparisons among the accuracy and diversity of CMB on dataset \textit{ML1M} when utilizing different $\lambda_2$.}
    \label{hy3}
    \centering
    \begin{adjustbox}{max width=\textwidth}
    \begin{tabular}{lcc|cc|cc|cc|cc|cc}
    \toprule
    \multirow{2}{*}{Metrics}
    & \multicolumn{2}{c}{Recall@K}
    & \multicolumn{2}{c}{NDCG@K}
    & \multicolumn{2}{c}{$\alpha$-nDCG@K}
    & \multicolumn{2}{c}{SC@K}
    & \multicolumn{2}{c}{PC@K}
    & \multicolumn{2}{c}{ILAD@K} \\
    \cmidrule{2-13}
    & \multicolumn{1}{c}{K=5}
    & \multicolumn{1}{c}{K=10}
    & \multicolumn{1}{c}{K=5}
    & \multicolumn{1}{c}{K=10}
    & \multicolumn{1}{c}{K=5}
    & \multicolumn{1}{c}{K=10}
    & \multicolumn{1}{c}{K=5}
    & \multicolumn{1}{c}{K=10}
    & \multicolumn{1}{c}{K=5}
    & \multicolumn{1}{c}{K=10}
    & \multicolumn{1}{c}{K=5}
    & \multicolumn{1}{c}{K=10}
    \\
    \midrule
    $\lambda_2 = 0.0$ & 0.0838 & 0.1382 & 0.2753 & 0.2581 & 0.6490 & 0.7087 & 0.3807 & 0.5098 & \textbf{0.2471} & \textbf{0.3301} & 0.2056 & \textbf{0.2262} \\
    $\lambda_2 = 0.1$ & 0.0839 & 0.1387 & 0.2757 & 0.2586 & 0.6502 & 0.7096 & 0.3811 & 0.5099 & 0.2458 & 0.3295 & \textbf{0.2057} & 0.2260 \\
    $\lambda_2 = 0.3$ & 0.0842 & 0.1385 & 0.2756 & 0.2580 & \textbf{0.6522} & \textbf{0.7111} & 0.3820 & 0.5097 & 0.2446 & 0.3283 & 0.2055 & 0.2261 \\
    $\lambda_2 = 0.5$ & 0.0841 & \textbf{0.1392} & 0.2761 & 0.2589 & 0.6495 & 0.7093 & \textbf{0.3821} & \textbf{0.5110} & 0.2454 & 0.3278 & 0.2047 & 0.2254 \\
    $\lambda_2 = 0.7$ & 0.0843 & \textbf{0.1392} & 0.2764 & \textbf{0.2590} & 0.6502 & 0.7099 & 0.3809 & 0.5100 & 0.2461 & 0.3288 & 0.2040 & 0.2247 \\
    $\lambda_2 = 0.9$ & 0.0842 & 0.1385 & 0.2760 & 0.2585 & 0.6491 & 0.7088 & 0.3806 & 0.5088 & 0.2424 & 0.3295 & 0.2038 & 0.2247 \\
    $\lambda_2 = 1.0$ & \textbf{0.0845} & 0.1388 & \textbf{0.2767} & \textbf{0.2590} & 0.6467 & 0.7071 & 0.3797 & 0.5087 & 0.2421 & 0.3291 & 0.2042 & 0.2248 \\
    \bottomrule
    \end{tabular}
    \end{adjustbox}
    \end{table*}

\end{document}